\documentclass{aa5}
\usepackage{psfig,latexsym,longtable,lscape}

\def\ros{{\sl ROSAT}}

\def\HI{\hbox{H\,{\sc i}}}
\def\HII{\hbox{H\,{\sc ii}}}
\newcommand{\D}{$^\circ$}
\def\p0{\phantom{0}}
\newcommand\approxlt{\mbox{$^{<}\hspace{-0.24cm}_{\sim}$}}

\def\it{\sl}

\begin{document}

   \title{ROSAT X-ray sources in the field of the LMC
   \thanks{Tables 1 and 3 are only available in electronic form 
   at the CDS via anonymous ftp to cdsarc.u-strasbg.fr 
   (130.79.128.5)}}
   \subtitle{II.Statistics of background AGN and X-ray binaries}

   \author{P. Kahabka\inst{1}}

   \offprints{P. Kahabka, \email{pkahabka@astro.uni-bonn.de}}
 
   \institute{Sternwarte, Universit\"at Bonn 
              Auf dem H\"ugel 71, D--53121 Bonn, Germany
             }

   \date{Received 27 August 2001 / Accepted 22 March 2002}

\abstract{
About 200 X-ray sources from a sample of spectrally hard \ros\ {\sl PSPC} 
sources, given in the catalog of Haberl \& Pietsch (1999), and observed 
in a $\sim$60 square degree field of the LMC during several archival pointed 
observations with a wide range of exposure times have been reanalyzed. For 
these sources accurate count rates and hardness ratios have been recalculated.
In comparison to Haberl \& Pietsch (1999) we used merged data from all
available observations and we derived average source parameters by 
investigating each source individually. From a simulation powerlaw spectral 
tracks have been derived in the $H\!R1$ -- $H\!R2$ plane and $\sim$170 sources 
have been classified as background X-ray sources or as LMC X-ray binaries. 
80\% of the spectrally hard X-ray sources with more than 50 observed counts 
have been found to be consistent with background X-ray sources and 20\% with 
LMC X-ray binaries (53 sources with AGN and 15 with X-ray binaries). The
discovery of a new supersoft source RX~J0529.4-6713 at the southern \HI\ 
boundary of the supergiant shell LMC\,4 is reported. We find two new candidate
X-ray binary systems which are associated with the optical bar of the LMC and
additional candidate X-ray binaries which are associated with supergiant 
shells. 
}
\titlerunning{Statistics of background AGN and X-ray binaries}
\authorrunning{P. Kahabka}
\maketitle
%
%________________________________________________________________
\section{Introduction}

X-ray background sources are active galactic nuclei (AGN) and clusters of 
galaxies for which the integral number versus flux distribution (the 
$\log N - \log S$) has been derived for different fields in the sky 
(cf. Hasinger et al. 1998; Gilli et al. 1999; 
Gilli et al. 2001). In the direction of nearby galaxies additional bright 
and spectrally hard X-ray sources are discovered which are associated with 
X-ray binaries (cf. for M31 Supper et al. 2001 and for M33 Haberl \& Pietsch 
2001). In the direction of the Magellanic Clouds which cover an area of a 
few 10 square degrees a few 100 background X-ray sources have been detected 
in deep pointed observations (cf. Haberl \& Pietsch 1999a, Haberl et al. 
2000). The $\sim$20--30 pulsating X-ray sources discovered in the Magellanic 
Clouds are to a large fraction associated with Be-type X-ray binaries (e.g. 
Liu et al. 2000). In addition a few bright blackhole X-ray binaries (LMC~X-1 
and LMC~X-3) exist in the LMC. In an investigation of the \ros\ all-sky 
survey ({\sl RASS}) observations of a 13\D\ by 13\D\ field centered on the 
LMC, more than 500 X-ray sources have been found (Pietsch \& Kahabka  1993). 
The \ros\ all-sky survey covered a large field of the LMC area and was only 
in regions close to the elliptical pole (north east of the LMC disk) deep 
enough to allow a detailed analysis of X-ray sources. Making use of the 
{\sl RASS} faint source catalog (Voges et al. 2000) $\sim$1200 sources are
found within about the same investigated field. For this sample a limiting
flux of $1.3\ 10^{-13}\ {\rm erg}\ {\rm cm}^{-2}\ {\rm s}^{-1}$ is derived
assuming a galactic foreground absorbing column of 
$5\ 10^{\rm 20}\ {\rm cm}^{-2}$ and a count rate to flux conversion factor
for a powerlaw photon spectrum with $-\Gamma=2.0$.
A much deeper survey of the LMC field has been performed during pointed \ros\ 
{\sl PSPC} (and {\sl HRI}) observations. A catalog of \ros\ {\sl PSPC} X-ray 
sources in a 10\D\ by 10\D\ field of the LMC has been established by Haberl \&
Pietsch (1999a). The catalog comprises 758 X-ray sources and identifications 
are given for 144 sources. In addition time variability has been taken into 
account by Haberl \& Pietsch (1999b) to refine the classification of $\sim$15 
LMC X-ray sources. For this sample, derived in the deeper pointed observations 
of the LMC field, a limiting flux of $4\ 10^{-15}\ {\rm erg}\ {\rm cm}^{-2}\ 
{\rm s}^{-1}$ is derived. (Subsequently it is referred to the Haberl \& 
Pietsch 1999a and 1999b papers as HP99). A catalog of \ros\ {\sl HRI} X-ray 
sources in a 10\D\ by 10\D\ field of the LMC has been established by Sasaki, 
Haberl \& Pietsch (2000, hereafter SHP00). They found 397 X-ray sources of 
which 259 are new detections in addition to the \ros\ {\sl PSPC} X-ray sources
found by Haberl \& Pietsch. A fraction of the classifications of the \ros\ 
sources in the LMC field are not firm partly due to lacking optical 
identifications and partly due to X-ray characteristics which allow more than 
one source class. 

I reanalyzed a considerable fraction of the sample of classified and 
unclassified spectrally hard X-ray sources in the LMC field and given in 
the \ros\ {\sl PSPC} catalog of HP99. I combined (merged) the observational 
data of these sources which have been obtained when the source was at \ros\ 
{\sl PSPC} off-axis angles of up to 50\arcmin. In general the analysis has 
been restricted to observations where the source was at an off-axis angle 
of $\approxlt30$\arcmin. I give in Sect.\,2 the selection criteria for the 
sample of spectrally hard sources. I will derive in Sect.\,3 the basic 
properties of this reanalyzed sample. I especially derive count rates in the 
spectrally hard (0.5 -- 2.0~keV) and broad (0.1 -- 2.4~keV) band as well as 
the X-ray colors (hardness ratios $H\!R1$ and $H\!R2$). For few sources 
positions more accurate than given in HP99 are derived. 
In Sect.\,4 I derive from simulations tracks for powerlaw spectra in the 
$H\!R1$ -- $H\!R2$ plane. It is made use of these tracks to achieve a 
classification of the spectrally hard sources as AGN or as X-ray binaries. 
I also take into account the AGN sample for which X-ray spectral fitting has 
been performed in Kahabka et al. (2001, Paper\,I). In addition I constrain
in Sect.\,5 for the classified background X-ray sources the hydrogen absorbing 
column density due to LMC gas assuming constraints on the powerlaw photon 
index for AGN type spectra. For a sub-sample of candidate AGN and X-ray 
binaries I also derive constraints on the metallicity of the LMC gas. 
In Sect.\,6 the X-ray binary sample is discussed and the number of X-ray 
binaries derived for the LMC is compared with the number of X-ray binaries 
observed in the SMC. Finally, in Appendix~A the catalog of X-ray sources in 
the field of the supergiant shell LMC\,4 will be given.

\section{Selection of the AGN and X-ray binary sample}

The sample of spectrally hard sources has been taken from the catalog of 
HP99. It comprises the sources which have already been identified by HP99
with AGN, quasars and galaxies. This sample is largely the sample which
has been used in Paper\,I for X-ray spectral fitting. A few sources have less 
than $\sim$100 counts and no X-ray spectra have been fitted for these 
sources. In addition, sources classified as $[hard]$ by HP99 have been
investigated. Source and background spectra have been created and from these 
data hardness ratios have been determined. These hardness ratios are for 
the {\sl ROSAT} {\sl PSPC} defined as

\begin{equation}
  H\!R1 = \frac{(Hard-Soft)}{(Hard+Soft)}
\end{equation}

\begin{equation}
  H\!R2 = \frac{(Hard2-Hard1)}{(Hard1+Hard2)}
\end{equation}

They are determined from the counts in the standard {\sl ROSAT} {\sl PSPC} 
bands $Soft$=(channel 11-41, 0.1-0.4\,keV), $Hard$=(channel 52-201, 
0.5-2.0\,keV), $Hard1$=(channel 52-90, 0.5-0.9\,keV) 
and $Hard2$=(channel 91-201, 0.9-2.0\,keV). The location of a source in the 
$H\!R1$ -- $H\!R2$ plane has been used for a source classification: In case 
the source was located in the AGN band $-\Gamma = (2.0-3.0)$ it was classified 
as an AGN. The source has been classified as an X-ray binary in case it was 
``above'' the AGN band (in the regime of spectra with less steep powerlaw 
indices). Not all sources classified as $[hard]$ by HP99 are consistent with 
AGN or X-ray binaries (some sources are extended). 

This sample has been extended by taking sources from the HP99 catalog into 
account which have been selected using criteria on the hardness ratios $H\!R1$ 
and $H\!R2$ and in addition on the extent likelihood ratio $ML_{\rm ext}$ 
(the extent likelihood ratio equals $ML_{\rm ext} = -ln(P)$, with $P$ the 
probability that the measured photon distribution deviates from the instrument 
point-spread function). These criteria are: $H\!R1>0.5$, $H\!R2>0.1$, 
$H\!R2 - \delta H\!R2<0.4$ and $ML_{\rm ext}<30$ (for AGN) and $H\!R1>0.5$, 
$H\!R2 - \delta H\!R2>0.4$ and $ML_{\rm ext}<30$ (for X-ray binaries). Only 
sources which have been observed in the inner 20\arcmin\ of the detector have 
been used for the analysis. This selection may not be complete as AGN with 
very low absorbing columns (e.g. located in the outskirts of the LMC) are not 
necessarily considered. Therefore additional {\sl ROSAT} {\sl PSPC} sources 
have been investigated which were inside the 20\arcmin\ radius of the 
{\sl PSPC} detector and which had at least 100 counts in the broad spectral 
band. For these sources a source and a background spectrum have been created 
and from these data hardness ratios have been calculated. Taking the value of 
the galactic absorbing column and the low LMC absorbing column into account 
the location of these sources in the $H\!R1$ -- $H\!R2$ plane has been used 
to classify these sources. There were 141 sources which were classified as AGN 
based on this selection. The X-ray binary sample comprises 30 sources and
is discussed in more detail in Sect.\,6.

\section{Determination of the basic source parameters}

The basic source parameters are the coordinates, the count rate in the
spectrally broad and hard band, the spectral colors (hardness ratios 
$H\!R1$ and $H\!R2$) and the amplitude of time variability. The hardness 
ratios will be used in a later section for a source classification. The 
count rate is used to construct the $\log N - \log S$ relation. As 
the standard $\log N - log S$ uses the count rate in the hard band 
it is of importance that this count rate is made available. I note that 
usually only the count rate in the broad band is given in catalogs. 
Conversion from one to another is in principle possible (from simulations) 
but then one has to assume the proper spectral index which is a priori not 
known. I recalculated the count rates and the hardness ratios for a subsample 
using the merged data. The values which are given are the mean values 
integrated over all observations. I applied the {\sl EXSAS} spectral fitting 
task (cf. Zimmermann et al. 1994) for binning and instrument correction 
(vignetting and dead time). I have individually chosen a circular region 
for the background subtraction which was in general close to the source and
had the same size as the source circle. This procedure allowed precise 
values to be derived and the statistical errors to be minimized. It has to 
be noted that values for the hardness ratios $H\!R1$ and $H\!R2$ which are 
larger 
than 1.0 are possible in case the source has a negligible flux in the 
0.1--0.5~keV band and background subtraction results in a negative count 
rate for this band. Still consideration of the errors in the hardness ratios 
for such cases allows values for the hardness ratios which are consistent 
with 1.0 (in Fig.\,1, 3, 4, and 5 we present or consider sources with 
hardness ratios $H\!R1 > 1.0$ and $H\!R2 > 1.0$ in the following way:
We set $H\!R1 = 1.0$ and $H\!R2 = 1.0$ and we show the error bar towards the
minimum value).  

In general the source coordinates have not been recalculated and 
it will be referred to the coordinates given in HP99.

\begin{figure}[htbp]
  \centering{
  \vbox{\psfig{figure=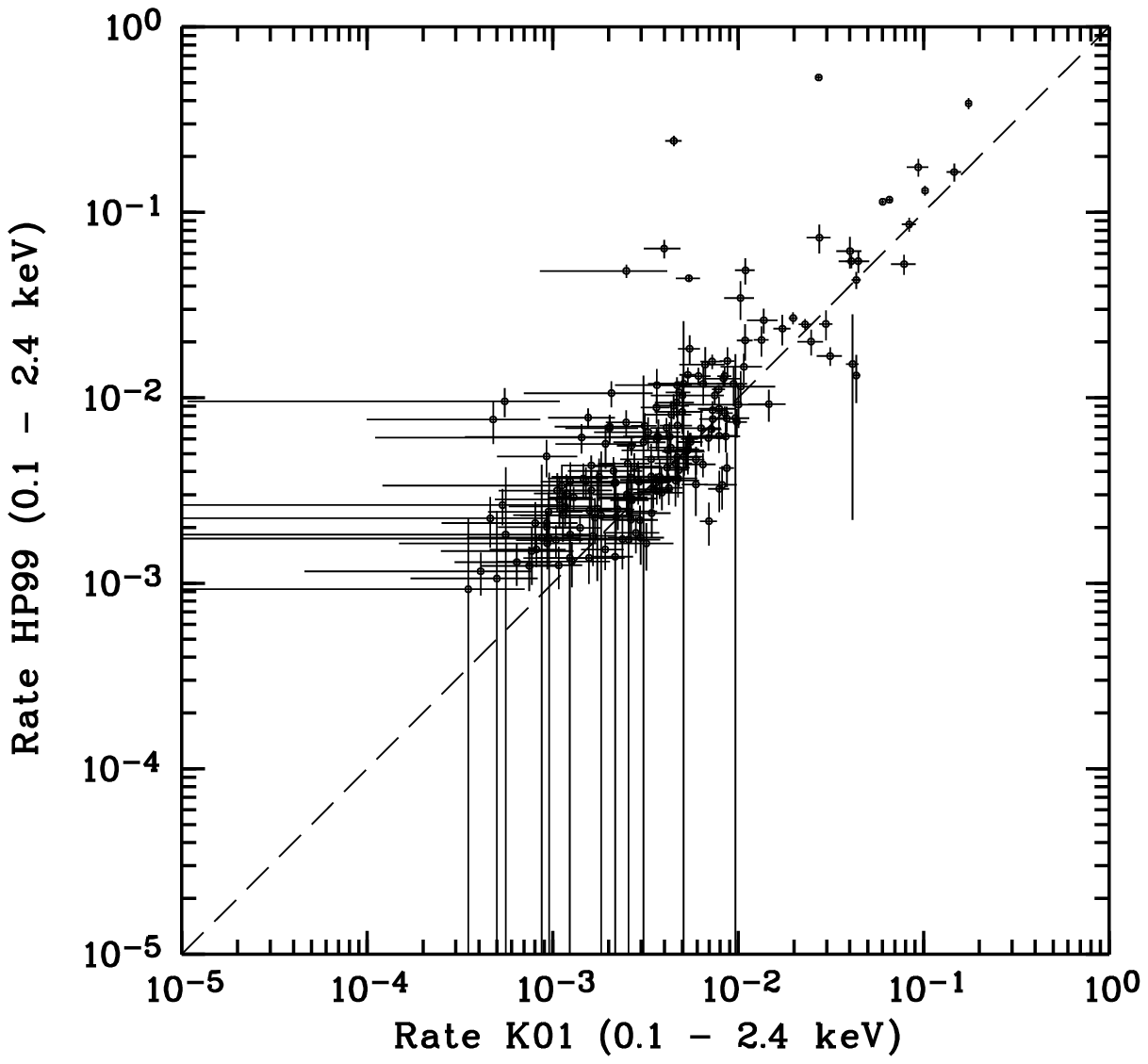,width=5.5cm,angle=0.0,%
  bbllx=3.5cm,bblly=1.5cm,bburx=16.5cm,bbury=13.5cm,clip=}}\par
  \vbox{\psfig{figure=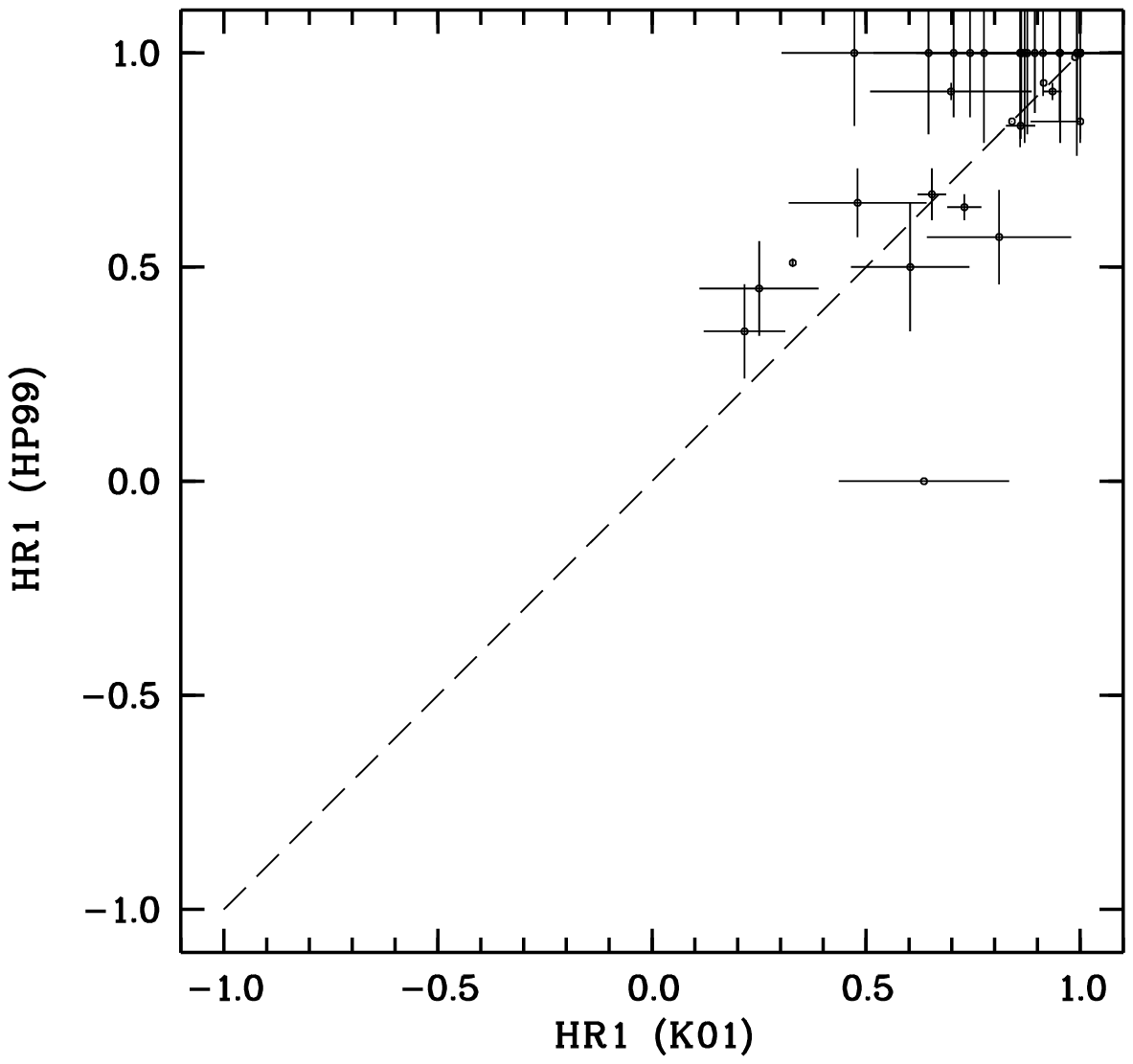,width=5.5cm,angle=0.0,%
  bbllx=3.5cm,bblly=1.5cm,bburx=16.5cm,bbury=13.5cm,clip=}}\par
  \vbox{\psfig{figure=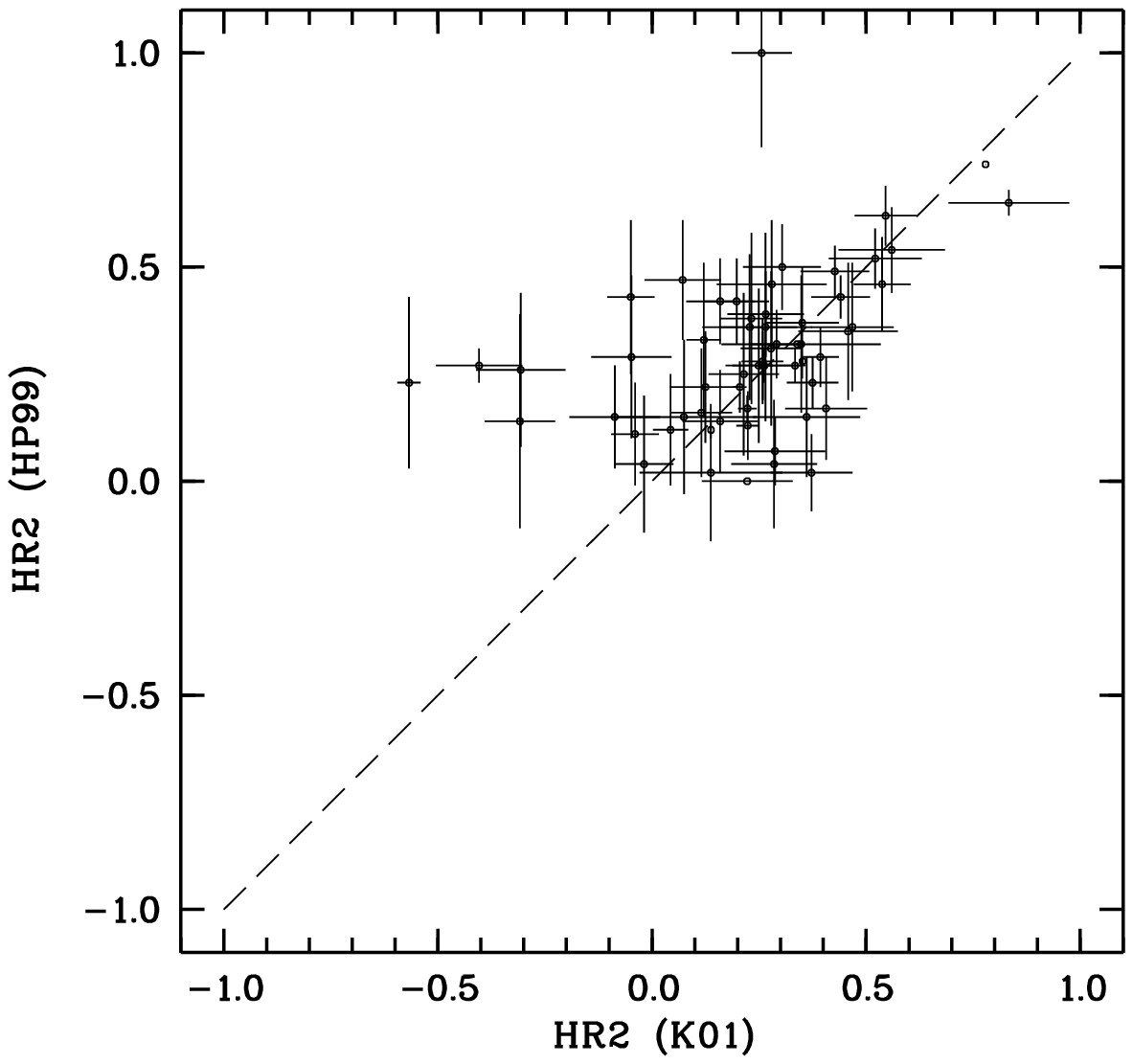,width=5.5cm,angle=0.0,%
  bbllx=3.5cm,bblly=1.5cm,bburx=16.5cm,bbury=13.5cm,clip=}}\par
            }
  \caption[]{Upper panel: Broad band (0.1 -- 2.4\,keV) {\sl PSPC} count 
  rates as derived in this work (K01, Tab.\,1) in comparison with the 
  {\sl PSPC} count rates derived by HP99. Middle panel: Soft hardness ratio 
  $H\!R1$ as derived in this work (K01) in comparison with $H\!R1$ derived 
  by HP99 (only sources with $\delta H\!R1<0.25$ have been plotted). 
  Lower panel: Hard hardness ratio $H\!R2$ as derived in this work (K01, 
  Tab.\,1) in comparison with $H\!R2$ derived by HP99 (only sources 
  with $\delta H\!R2<0.25$ have been plotted). For sources with 
  $H\!R1 > 1.0$ and $H\!R2 > 1.0$ a value of 1.0 is given and the error
  bar towards the minimum allowed value.}
  \label{ps:figrate}
\end{figure}

\subsection{Comparison with the source parameters given in HP99}

A different approach has been chosen than has been used by HP99 to derive 
source parameters (count rates and hardness ratios). I have derived mean 
parameters from the merged data of all available observations with integration
times of at least 1000 seconds. This, in general, reduced the statistical 
errors. In addition spectra (source and source plus background) 
have been accumulated for each individual source and the standard instrument 
corrections have been applied. For time variable sources average values have 
been derived in the analysis. It is expected to derive more precise values for 
the source count rate and the X-ray colors (hardness ratios $H\!R1$ and 
$H\!R2$) than given in HP99. In Fig.~1 (upper panel) the broad band 
(0.1 -- 2.4 keV) \ros\ {\sl PSPC} count rates as derived by HP99 and in this 
analysis are compared. The count rate derived by these two methods follows 
the same trend but individual rates scatter. In addition it is found that
the count rates derived by HP99 are systematically larger (by about a factor 
of 1.5) than the count rates derived by K01. Such a systematic difference
may be explained by the different method which has been applied to determine
the source counts. In addition HP99 determined the count rate of a source for
the observation (in case multiple observations exist) in which the derived
position was most accurate. In case of time variable sources such an
approach can bias the count rate to values which are systematically larger 
than the mean value averaged over many observations. The importance of this 
effect was checked for two bright AGN in the field of the LMC 
(RX~J0524.0-7011 and RX~J0503.1-6634) which were found to be variable with 
a timescale of a few hundred days to few years. It was found that for these 
two AGN the count rate averaged over observations spread over a few years 
was a factor of 3 and 1.6 respectively smaller than the count rate given in 
the catalog of HP99. In addition it is found that the count rates of 17 
sources deviate more than $5\sigma$ and of 9 sources more than $10\sigma$.
7 of the 9 sources are X-ray binaries which show large variability in the
count rate with time. In the catalog presented here count rates averaged over 
all observations are given while in HP99 the count rate for the observation 
is given where the source position has been determined (the largest deviation 
with a few hundred sigma is found for LMC~X-4, for this source a low count
rate is given in the catalog of HP99). The remaining two sources are the AGN 
RX~J0524.0-7011 and RX~J0503.1-6634 which are variable in time (see discussion
above).

In this work mainly spectrally hard sources (candidates for AGN and X-ray 
binaries) have been selected which have hardness ratios $H\!R1$ and $H\!R2$ 
$>0.0$. In Fig.~{\ref{ps:figrate} (middle and lower panel) the comparison 
between the hardness ratios $H\!R1$ and $H\!R2$ is shown as derived by both 
methods.for sources with precise hardness ratios ($\delta H\!R1 < 0.25$). 
There is some scatter in these values which may be explained by the 
different methods which have been applied. A few sources were found in this 
analysis to be softer than given in the catalog of HP99 (cf. the sources in 
the lower panel of Fig.\,1 which are found at $H\!R2$ values $<$0). These 
sources were not classified as AGN, cf. Tab.\,1). In the further analysis 
the values for the count rate and hardness ratios derived in this work have 
been used.}

\section{Source classification and parameter determination in the
hardness ratio plane}

In Paper\,I the technique of X-ray spectral fitting has been applied to a 
sample of 26 background AGN in the field of the LMC. These AGN were taken 
from the {\sl ROSAT} {\sl PSPC} catalog of LMC X-ray sources of HP99 and 
candidate AGN selected here. 

A different approach to constrain the spectral parameters is to use the 
hardness ratios $H\!R1$ and $H\!R2$.
These hardness ratios are commonly available for {\sl ROSAT} {\sl PSPC}
X-ray sources published e.g. in catalogs. There are two techniques to
calculate these hardness ratios. The first is to fit the {\sl ROSAT}
{\sl PSPC} point-spread function to data binned spatially in the
standard energy bands and to determine the background from a spline-fit 
applied to these binned data. Such a procedure can be applied to a
large sample of sources detected in the field of view of a {\sl PSPC}
observation. The second technique is to determine the source and the
background counts in the standard energy bands from observational data 
which have been corrected for the {\sl ROSAT} {\sl PSPC} instrument functions. 
The background counts have been derived from a spatial region close to the 
source.

In order to constrain the spectral parameters of a source the observed
hardness ratios $H\!R1$ and $H\!R2$ have been compared with the hardness
ratios determined from simulations. Such a comparison was performed with a
$H\!R1$ -- $H\!R2$ grid derived from such simulations. Free parameters to be
varied in the simulations were the powerlaw photon index $\Gamma$ of the 
adopted photon spectrum, the hydrogen absorbing column density $N_{\rm H}$, 
and the metallicity $X$. The redshift $z$ has not been varied in the 
simulations. It has always been set to $z=0$.

\begin{figure} 
  \centering{
  \vbox{\psfig{figure=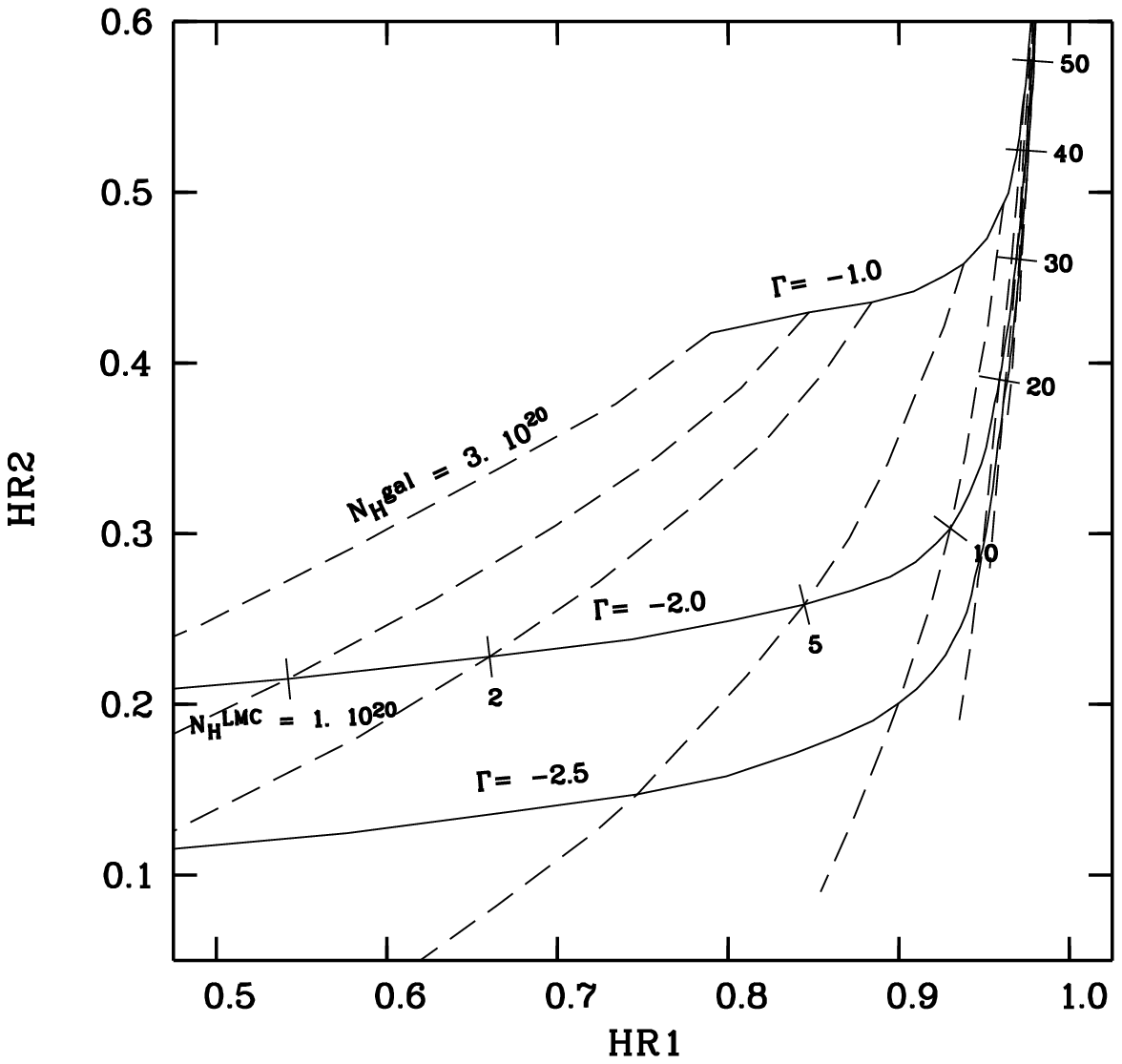,width=6.0cm,angle=0.0,%
  bbllx=4.3cm,bblly=1.5cm,bburx=16.3cm,bbury=13.5cm,clip=}}\par
  \vbox{\psfig{figure=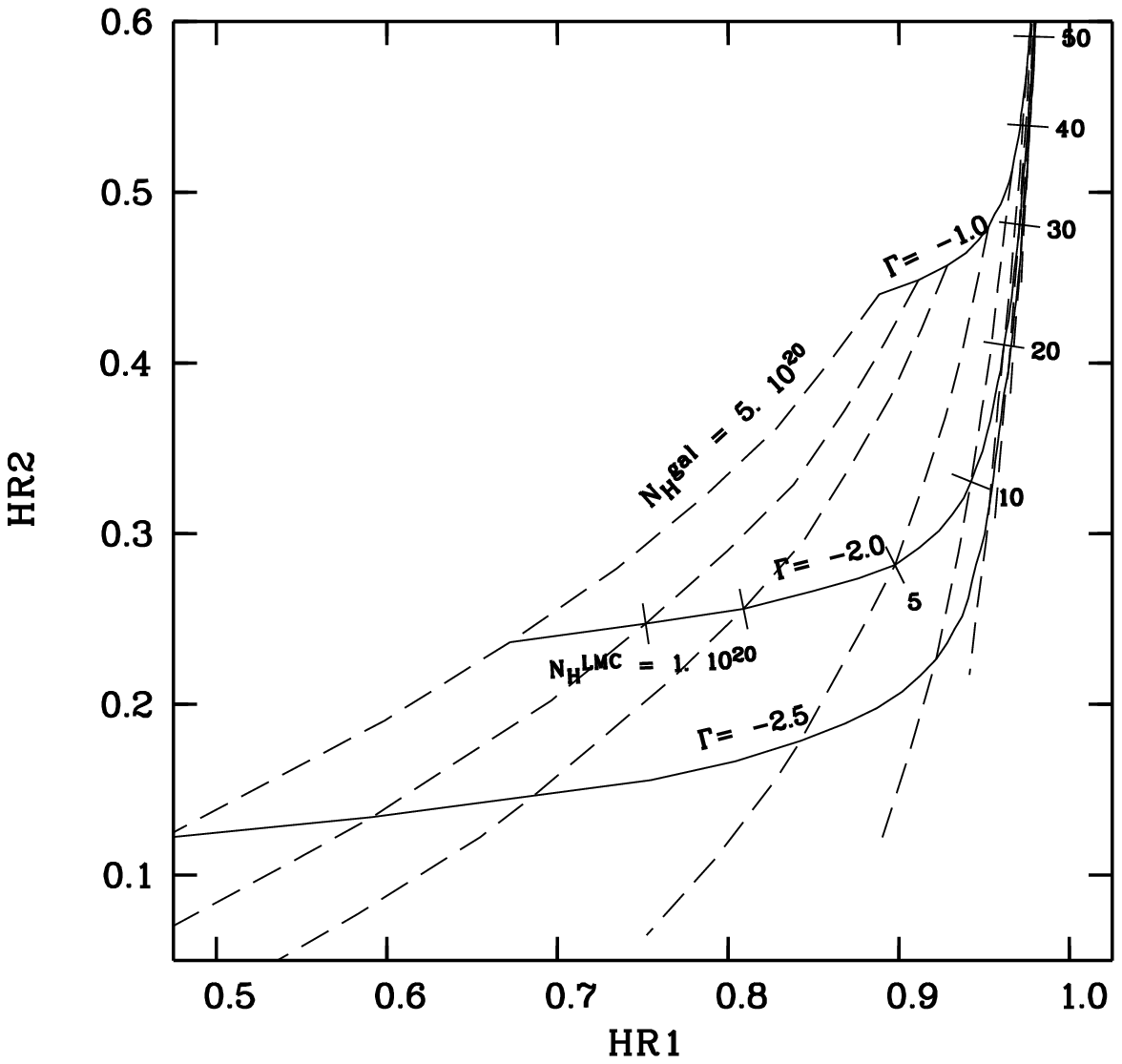,width=6.0cm,angle=0.0,%
  bbllx=4.3cm,bblly=1.5cm,bburx=16.3cm,bbury=13.5cm,clip=}}\par
            }
  \caption[]{Enlarged $H\!R1$ -- $H\!R2$ grid calculated for background 
  X-ray sources (AGN) and X-ray binaries, showing the effect of foreground 
  absorption. The full lines give $\Gamma$ isolines, the dashed lines 
  represent $N_{\rm H}$ isolines. The band for AGN type spectra is shown 
  which is defined by $-\Gamma$ = (2.0:2.5). The $N_{\rm H}$ isolines have 
  been labelled at the intersection with the $-\Gamma$ = 2.0 line with 
  $N_{\rm H}^{\rm LMC}$ in units of $10^{20}$ cm$^{-2}$. The leftmost (long 
  dashed) line is for zero LMC gas. The calculations include foreground 
  galactic gas with $N_{\rm H}^{\rm gal} = 3\,10^{20}\,{\rm cm}^{-2}$
  (upper panel) and $5\,10^{20}\,{\rm cm}^{-2}$ (lower panel), while for 
  the LMC gas a mean metallicity of $-0.3$ dex is assumed.}
  \label{ps:figsim1}
\end{figure}

In the case of the LMC background AGN, it is expected that the parameters can
be confined to well defined ranges. For example in recent work it has been 
found that AGN have canonical powerlaw indices which can be confined to 
a narrow range $-\Gamma = 2.0$ to $2.5$ for the {\sl ROSAT} {\sl PSPC} 
(cf Brinkmann et al. 2000). These canonical values may still somewhat depend 
on the chosen energy range (instrument). Also the mean metallicity of the LMC 
gas is quite well constrained from observational work (e.g. Dopita \& Russell
1992). These facts, in principle, allow the hydrogen absorbing column density 
towards an AGN from simulations to be determined. The redshift of the AGN 
has only a minor effect on the simulated spectra for the expected redshift 
range covered by the AGN sample (cf. Comastri et al. 1995). For the hydrogen 
column density $N_{\rm H}$ the model must account for the galactic 
contribution as well as the LMC contribution. These two components are assumed
to have different metallicities and these models will be termed $hybrid$ 
models. AGN can also show intrinsic absorption (cf. Comastri et al. 1995). 
But most of the intrinsically absorbed AGN will not be detected in the 
{\sl ROSAT} band as the value of the absorbing column is large and the fluxes 
are low.

In Fig.~{\ref{ps:figsim1}} I give the $H\!R1$ -- $H\!R2$ grid for a hybrid
$N_{\rm H}$ model. For the first component with galactic metallicity a value 
of $N_{\rm H}^{\rm gal}$ of $3$ and $5\times 10^{20}\ {\rm cm}^{-2}$ has been
used, for the second component mean LMC metallicities ($X = -0.3$\,dex with
respect to galactic interstellar absorption abundances, Morrison \& McCammon
1983) which range from $10^{20}\ {\rm cm}^{-2}$ to $10^{22}\ {\rm cm}^{-2}$ 
in steps of $10^{20}\ {\rm cm}^{-2}$. In this classification scheme X-ray 
binaries cover $-\Gamma$ values of $\sim$(0.5--1.6) and AGN cover $-\Gamma$ 
values of $\sim$(1.8--3.0).

The range chosen for the powerlaw photon index $\Gamma$ is the range which 
is presently considered to be the most reliable in the {\sl ROSAT} {\sl PSPC} 
band and is considered to be the canonical band. This range of $\Gamma$ values
agrees with the range of $\Gamma$ values required in Paper\,I to classify 
AGN from the result of X-ray spectral fitting.

I have derived the values for the hardness ratios and the errors in the 
hardness ratios making use of the same source and background regions 
as chosen for X-ray spectral fitting in Paper\,I. The spectral data have been
binned in the standard energy bands used in the hardness ratio definition and 
the data have been corrected using the {\sl EXSAS} correction package 
(Zimmermann et al. 1994). 

In Tab.\,1 the catalog of the reanalyzed spectrally hard X-ray sources as 
taken from the catalog of HP99 is given. One additional source, 
RX~J0536.9-6913, is contained in the catalog which is not contained in the 
catalog of HP99 but which has been investigated in Paper\,I and found to be 
consistent with an absorbed AGN. In the table first the sources classified 
as X-ray binaries are given, then the AGN, and at the end of the table a 
few sources classified as SNR or foreground stars are given. For the last 
11 sources in the table no classification is given. In Column~(1) of Tab.\,1 
the \ros\ name is given, in Column~(2) and (3) the source index from the same 
catalog and the catalog of Sasaki, Haberl, \& Pietsch (2000), in Column~(4) 
and (5) the count rate of the broad (0.1 -- 2.4~keV) and hard (0.5 -- 2.0~keV) 
band, in Column~(6) and (7) the hardness ratios $H\!R1$ and $H\!R2$ including 
$1\sigma$ errors, in Column~(8) and (9) the column density of the galactic 
and LMC \HI\ derived from 21-cm {\sl Parkes} data (Br\"uns et al. 2001), in 
Column~(10) the LMC column density derived from the hardness ratio analysis 
from this work, in Column~(11) the source classification, and in Column~(12) 
references and notes to individual sources.

\subsection{Source classification in the $H\!R1$ -- $H\!R2$ plane}

\begin{figure} 
  \centering{
  \vbox{\psfig{figure=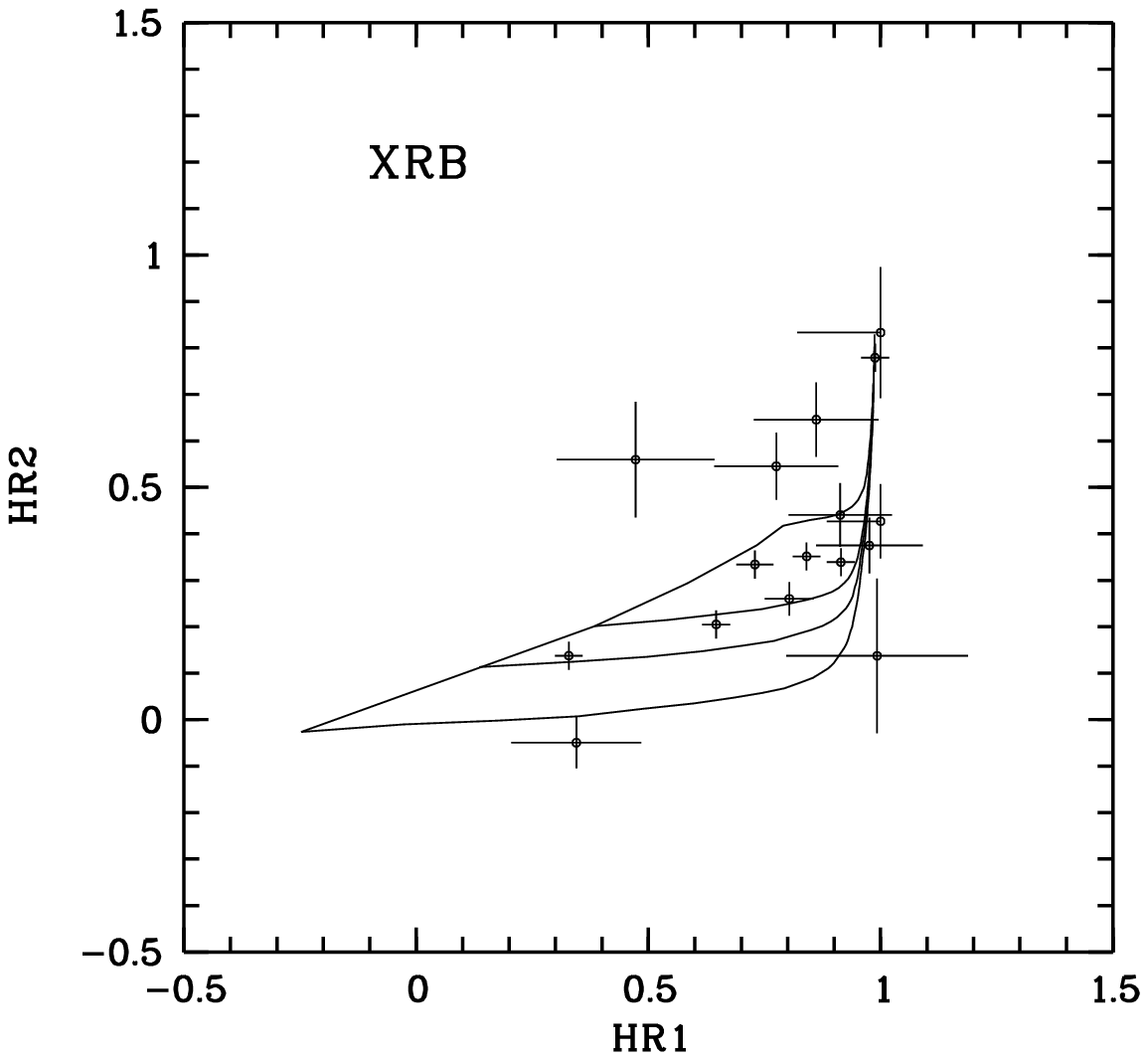,width=5.5cm,angle=0.0,%
  bbllx=2.0cm,bblly=1.5cm,bburx=14.0cm,bbury=13.0cm,clip=}}\par
  \vbox{\psfig{figure=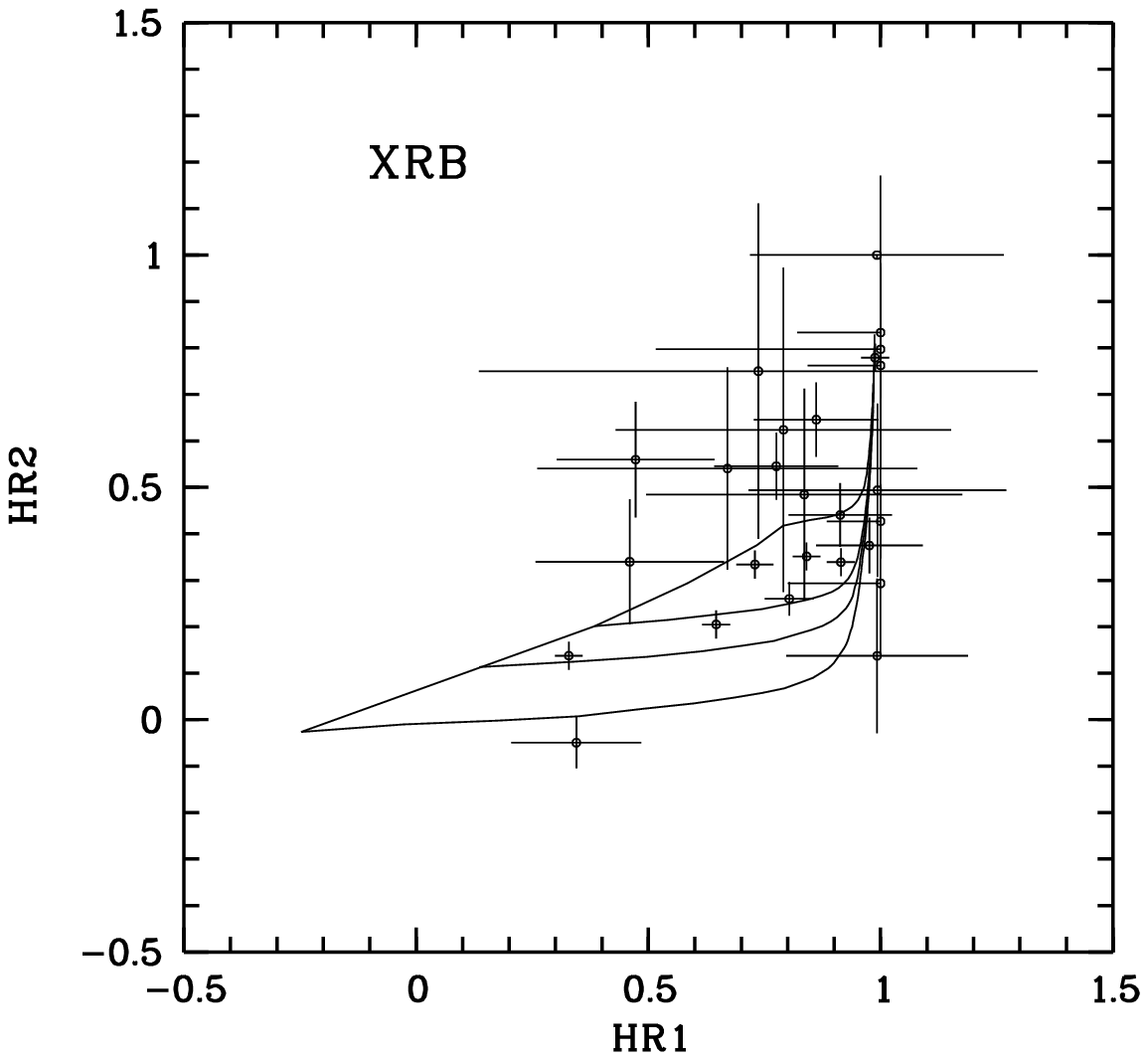,width=5.5cm,angle=0.0,%
  bbllx=2.0cm,bblly=1.5cm,bburx=14.0cm,bbury=13.0cm,clip=}}\par
  \vbox{\psfig{figure=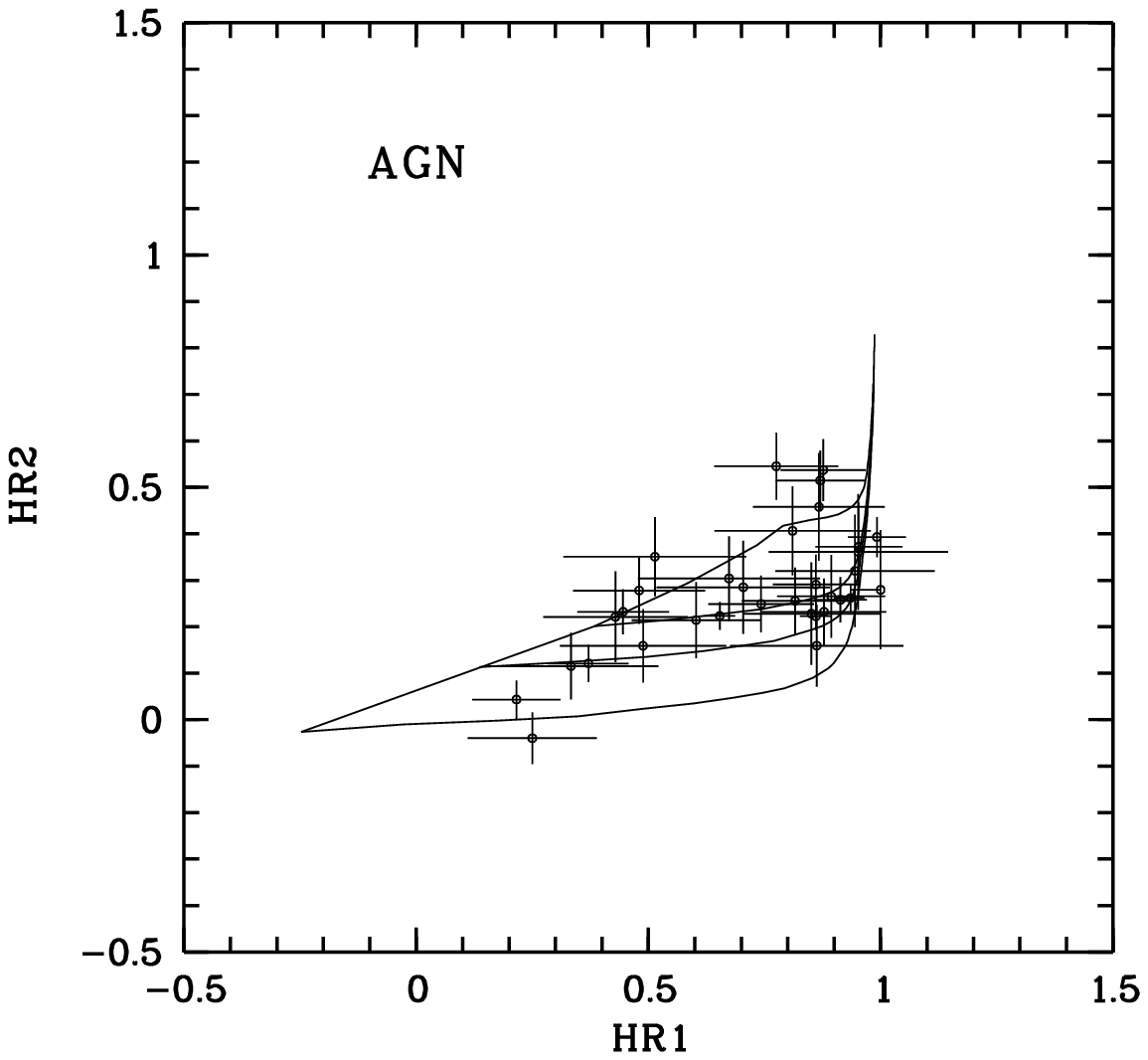,width=5.5cm,angle=0.0,%
  bbllx=2.0cm,bblly=1.5cm,bburx=14.0cm,bbury=13.0cm,clip=}}\par
  \vbox{\psfig{figure=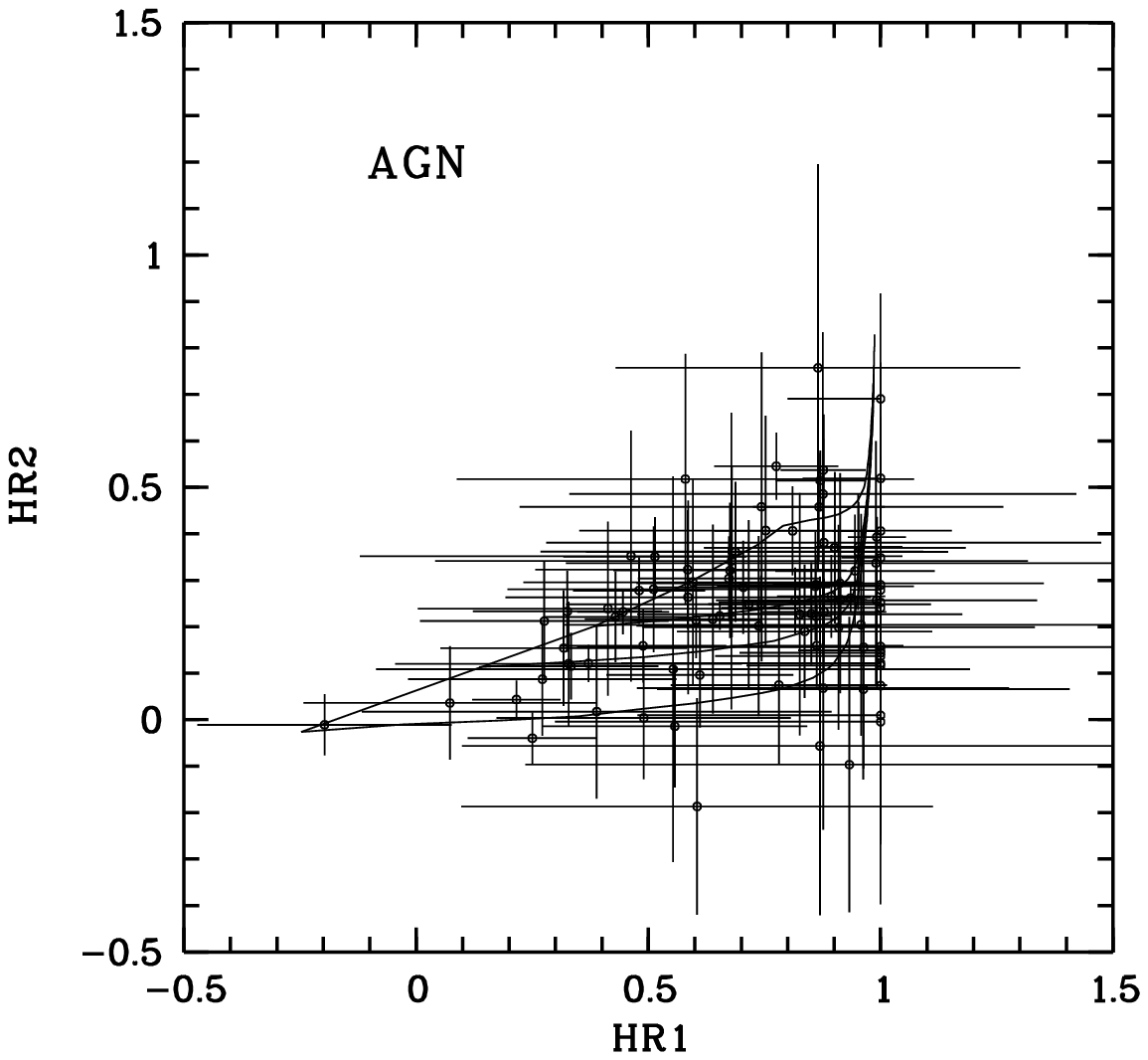,width=5.5cm,angle=0.0,%
  bbllx=2.0cm,bblly=1.5cm,bburx=14.0cm,bbury=13.0cm,clip=}}\par
            }
  \caption[]{Sources from Tab.\,1 classified as X-ray binaries (XRB, upper 
  two panels) and background AGN (lower two panels). First and third panel: 
  Sources with $\delta H\!R1 \le 0.2$ and $\delta H\!R2\ \le 0.2$ are given. 
  Second and fourth panel: Sources with $\delta H\!R1 \le 0.85$ and 
  $\delta H\!R2\ \le 0.85$ are given. Also shown are the simulated 
  (powerlaw slope $-1.0$,  $-2.0$, $-2.5$, and $-3.0$) tracks for hydrogen 
  columns in excess of galactic hydrogen columns ($N_{\rm H}^{LMC} > 0.0$) 
  and assuming a value for the galactic hydrogen column density of 
  $N_{\rm H}^{\rm gal} = 3\ 10^{20}\ {\rm cm^{-2}}$. For sources with
  $H\!R1 > 1.0$ and $H\!R2 > 1.0$ a value of 1.0 is given and the error
  bar towards the minimum allowed value.}
  \label{ps:figxrb}
\end{figure}

I made simulations in which I varied the powerlaw photon index $\Gamma$ of 
the source spectrum and the LMC hydrogen column density $N_{\rm H}^{\rm LMC}$ 
assuming reduced metallicities which are expressed with the logarithmic 
decrement $X$. Different values have been assumed for the powerlaw photon 
index $\Gamma$ for X-ray binaries and AGN ($-\Gamma = 0.5$ to $1.6$ for 
X-ray binaries and $-\Gamma = 1.8$ to $3.0$ for AGN).

\begin{figure} 
  \centering{
  \vbox{\psfig{figure=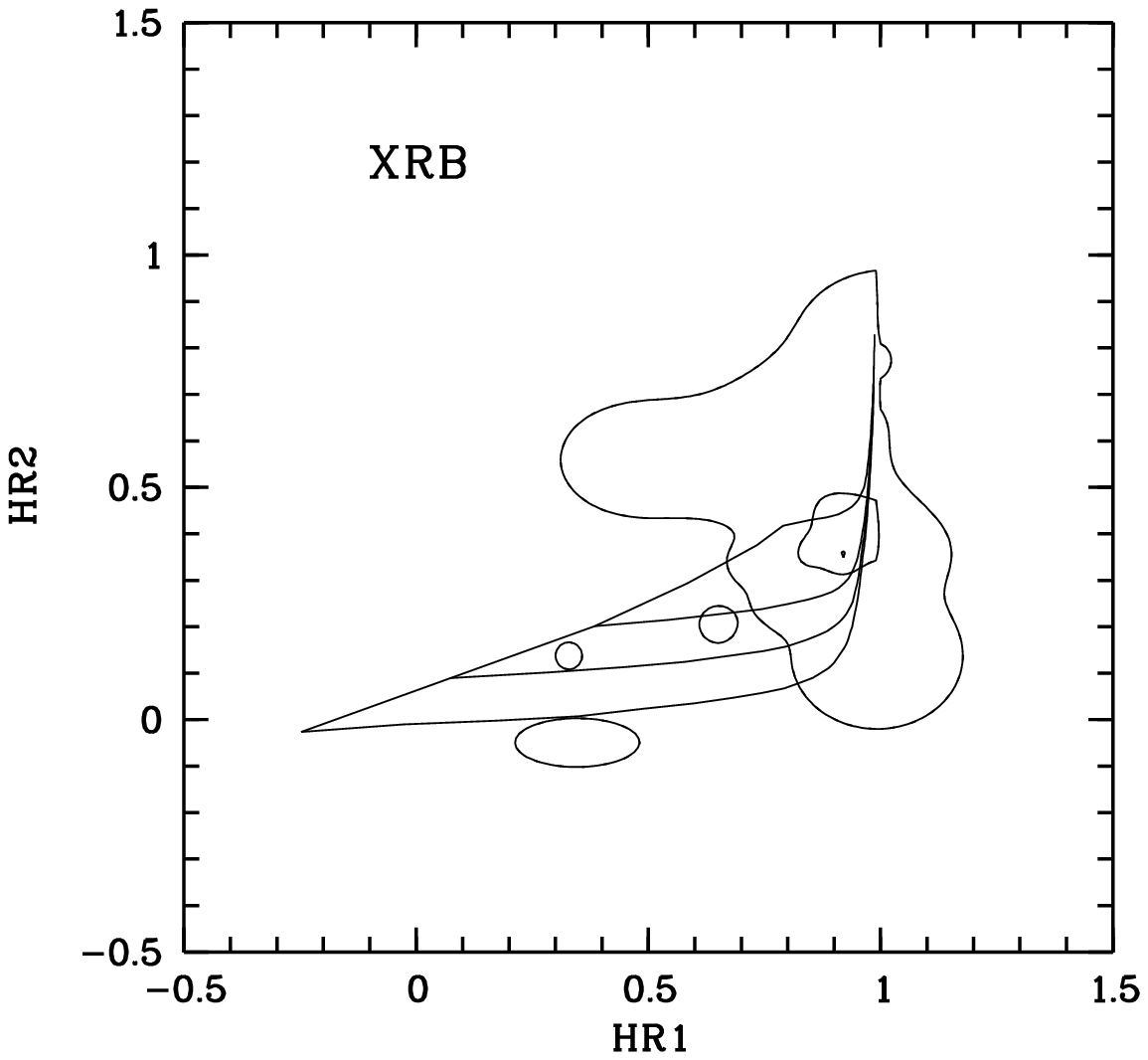,width=5.0cm,angle=0.0,%
  bbllx=2.0cm,bblly=1.5cm,bburx=14.0cm,bbury=13.0cm,clip=}}\par
  \vbox{\psfig{figure=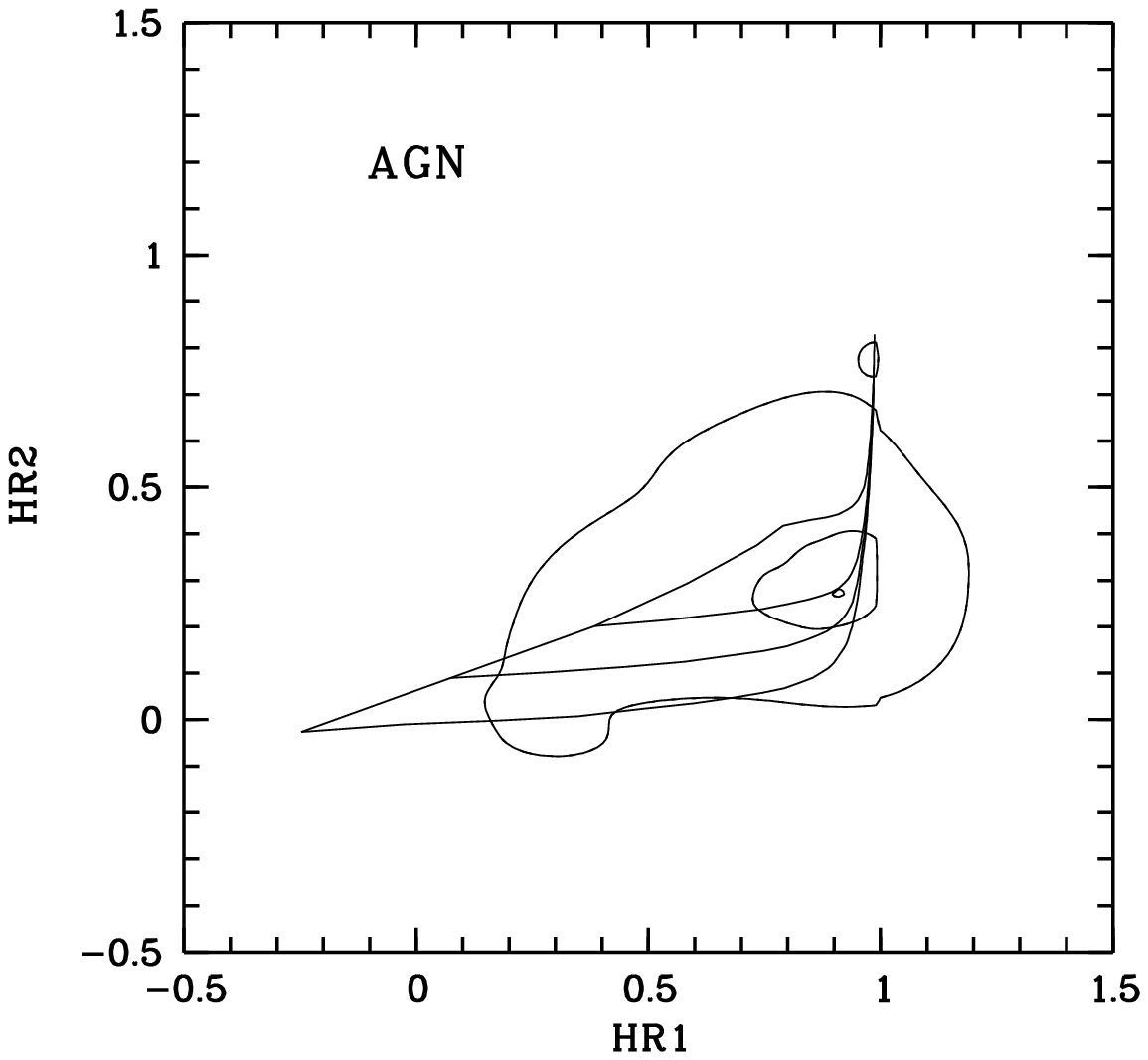,width=5.0cm,angle=0.0,%
  bbllx=2.0cm,bblly=1.5cm,bburx=14.0cm,bbury=13.0cm,clip=}}\par
            }
  \caption[]{Probability distribution of $H\!R1$ and $H\!R2$ values in the 
  $H\!R1$ -- $H\!R2$ plane. The 68\%, 95.4\% and 99.7\% confidence contours 
  are given for sources with $\delta H\!R1\le0.2$ and $\delta H\!R2\le0.2$
  from Tab.\,1 classified as X-ray binaries (XRB, upper panel) and as AGN 
  (lower panel). Also given are the tracks with constant powerlaw photon 
  index $-\Gamma = 1.0,2.0,2.5$ and $3.0$.}
  \label{ps:figmdm}
\end{figure}

From the location of a source in the $H\!R1$ -- $H\!R2$ plane a tentative 
source classification has been made. Sources which have hardness ratios
which coincide with the range of $\Gamma$-tracks for X-ray binaries or 
AGN have been classified accordingly. In addition the galactic and LMC
column at the location of a source has been used for a source classification.
AGN are supposed to be seen through the galactic and total LMC absorbing 
column and X-ray binaries are seen through at least the galactic and at 
most the total LMC absorbing column. Of course there may be sources which 
have spectral properties which deviate from the standard values and the 
classification may not be unique. Especially the similarity between the 
spectral properties of the low-mass X-ray binary LMC\,X-2 and AGN is striking 
(LMC\,X-2 is located in the regime of AGN type spectra). A few of the sources 
which have not been classified as candidate X-ray binaries by HP99 and which 
show time variability in X-rays are located in the AGN regime and could also 
be time variable AGN. The source RX~J0532.7-6926 (with number 914 in the 
catalog of HP99) has been classified as a LMXB by Haberl \& Pietsch 1999b
from a time variability analysis. From X-ray spectral fitting follows that 
this source has a very steep spectrum ($-\Gamma$$\sim$3.0) and could also
be a time variable AGN. We did not include this source in the further
analysis in the class of X-ray binaries.

In the color -- color diagram ($H\!R1$ -- $H\!R2$ plane), 
Fig.~{\ref{ps:figxrb}}, I show the location of the sources classified as 
X-ray binaries and as background AGN as given in Tab.\,1. It is obvious 
that X-ray binaries and background AGN cover in general different areas 
in this diagram as expected due to the different steepness of their spectral 
slopes. Background AGN have steeper slopes and are found in regimes of 
lower values for $H\!R2$ than X-ray binaries (cf. Fig.~{\ref{ps:figmdm}}) 
although there is some region of overlap (some X-ray binaries have as steep 
X-ray spectra as AGN). Due to absorption by galactic gas with column 
densities in the range $\sim$$(3-7)\times 10^{20}\ {\rm cm}^{-2}$ the value 
of $H\!R1$ does not extend to values $H\!R1\approxlt0$.

\begin{figure} 
  \centering{
  \vbox{\psfig{figure=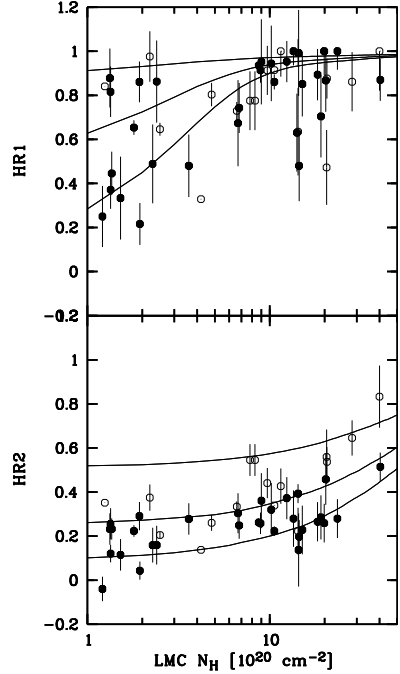,width=5.5cm,angle=0.0,%
  bbllx=1.0cm,bblly=0.7cm,bburx=9.0cm,bbury=14.2cm,clip=}}\par
            }
  \caption[]{Sources from Tab.\,1 classified as background AGN or X-ray 
  binaries and in the LMC $N_{\rm H}$ -- $H\!R1$ and $H\!R2$ plane. Tracks 
  for three different powerlaw photon indices $-\Gamma$ = 0.5, 1.8 and 2.5 
  (upper to lower curves) are given. The sample of 49 AGN (filled circle) 
  and X-ray binaries (open circle) is shown. Sources with accurate hardness 
  ratio values $\delta H\!R1 \le 0.20$ and $\delta H\!R2\ \le 0.20$ have 
  been used. The LMC $N_{\rm H}$ has been derived local to the X-ray source 
  from 21-cm {\sl Parkes} data and making use of the galactic foreground 
  $N_{\rm H}$ derived from 21-cm {\sl Parkes} data. Two AGN with 
  $N_{\rm H}^{\rm LMC} < 10^{20}\ {\rm cm}^{-2}$ are not shown in the 
  figure.} 
  \label{ps:classnhhr}
\end{figure}

I can obtain information about the total hydrogen column density due to 
LMC gas from the source shape. Due to the dependence of the \ros\ {\sl PSPC} 
instrument point-spread function on the energy (the point-spread-function 
becomes narrower with increasing energy) AGN seen through high LMC columns 
appear sharp and pointlike while AGN seen through low LMC columns have broader
images. This fact helps to confirm large LMC columns derived in the direction 
of background AGN. But some classified AGN may be X-ray galaxies and 
intrinsically extended so this argument is not a perfect one.

A new source has been detected in the merged \ros\ {\sl PSPC} pointings
(see Tab.\,2) which is not contained in the catalog of HP99.\footnote{It 
is a new supersoft source RX~J0529.4-6713 at the southern boundary of the 
supergiant shell LMC\,4. This source is very close to another source 
RX~J0529.7-6713 which is contained in the HP99 catalog. Due to the closeness 
of both sources (the other source has been classified as an AGN) the source 
detection algorithm may have excluded this source. Spectral fitting using a 
blackbody spectral shape gives a temperature of 40~eV and a luminosity of 
$\sim$$2\ 10^{36}\ {\rm erg}\ {\rm s^{-1}}$ if a galactic absorbing column 
density of $5.6\ 10^{20}\ {\rm cm^{-2}}$ and a LMC column density of 
$4.7\ 10^{20}\ {\rm cm^{-2}}$ is used. The spectrum is consistent with such 
absorption which is in favor for a supersoft source in the LMC,
the southern region of LMC\,4. The source appears to be variable in time. 
The large luminosity of the source would argue against a conventional CV 
nature of the source as for CVs luminosities do not exceed 
$\sim$$10^{33}\ {\rm erg}\ {\rm s^{-1}}$. It thus could be a white dwarf 
with steady nuclear burning. Inside the 11\arcsec error circle no optical 
counterpart is seen in digital sky survey images (blue and red plates). There 
are 2 to 3 optical stellar objects at the periphery of the error circle.}

A second source is given which is not included in the catalog of HP99, the 
heavily absorbed background source RX~J0532.0-6919 in the 30~Dor complex
which coincides in position with the radio source MDM\,65 of Marx, Dickey, \&
Mebold (1997). In addition two sources are given for which significantly 
improved positions (compared to those given in HP99) were derived.

Column\,1 of the catalog of Tab.\,2 gives the catalog index, Column\,2 the 
{\sl ROSAT} source name, Column\,3 and 4 the source position, the right 
ascension (RA) and declination (Dec) for the epoch J2000 with the 90\% 
confidence positional uncertainty (Column\,5), the likelihood of existence 
$L_{\rm exi} = -ln(P)$ (Column\,6), with $P$ the probability that the detected 
source is due to excess counts measured above a mean local background. For 
the first source, RX~J0529.4-6713, in addition to the coordinates the 
values of the hardness ratios $H\!R1$ and $H\!R2$ are given. The source 
parameters have been determined by applying the maximum likelihood source 
detection task to the merged data in the field of the corresponding source.

\addtocounter{table}{1}
\begin{table}[htbp]
  \caption[]{New X-ray sources detected in merged \ros\ {\sl PSPC} 
  observations and X-ray sources from the HP\,99 catalog for which 
  the positions have been improved.}
  \begin{flushleft}
  \begin{tabular}{cccccc}
  \hline
  \noalign{\smallskip}
(1)    &(2)     &(3)      &(4)      &(5)  &(6)                         \\
Source & Source & RA      & Dec     &$\rm P_{\rm e}$ & $L_{\rm exi}$   \\
No.    & Name   & (J2000) & (J2000) &                &                 \\
       & RX~J   & \p0h~\p0m~\p0s\p0~ & \p0\D\p0\p0$\arcmin$~\p0$\arcsec$  & ($\arcsec$) \\
  \noalign{\smallskip}
  \hline
  \noalign{\smallskip}
   1   & 0529.4-6713 & 05 29 25.8 & -67 13 24  & 11  &  91 \\
   2   & 0529.7-6713 & 05 29 47.0 & -67 13 50  & 11  &  10 \\
   3   & 0536.9-6913 & 05 36 57.9 & -69 13 29  & 17  & 125 \\
   4   & 0553.2-7144 & 05 53 13.4 & -71 44 03  & 25  &  11 \\
  \noalign{\smallskip}
  \hline
  \noalign{\smallskip}
  \end{tabular}
  \end{flushleft}
  \label{tab:newsources}
  Notes on sources: Source~1: $H\!R1 = -0.849\pm0.142$, 
  $H\!R2 = 0.474\pm0.020$, close to source~2; 
  Source~2: HP\,494, close to source~1; 
  Source~3: 180~ksec exposure, MDM65, see also Tab.\,1; 
  Source~4: HP\,1303, 17\arcmin off-axis.
  \label{tab:newsources}
\end{table}

In Fig.~{\ref{ps:classnhhr}} I show the location of 49 X-ray sources 
from Tab.\,1 with accurate values for the hardness ratios 
$\delta H\!R1 \le 0.20$ and $\delta H\!R2 \le 0.20$ and which have been 
classified either as background AGN or as X-ray binaries in the LMC 
$N_{\rm H}$ -- $H\!R1$ and $H\!R2$ plane respectively. The two AGN HP\,37 
and HP\,352 which have $N_{\rm H}^{\rm LMC} < 10^{20}\ {\rm cm}^{-2}$ are 
not shown in this figure. The source HP\,414 has not been included in the 
sample as it may be a foreground object and also the source HP\,914 has 
not been included as it may either be a LMXB or an AGN. I also give tracks 
for powerlaw photon indices $-\Gamma$ = 0.5, 1.8 and 2.5. Sources classified 
as XRB are preferrentially found in the $-\Gamma$ = 0.5 to 1.8 band while 
sources classified as AGN are preferentially found in the $-\Gamma$ = 1.8 
to 2.5 band. There are a few exceptions, e.g. the XRB LMC~X-4 has a steep 
powerlaw photon index and is outside the $-\Gamma$ = 0.5 to 1.8 band. The 
source classification has been made using the LMC $N_{\rm H}$ -- $H\!R2$ 
diagram (in the LMC $N_{\rm H}$ -- $H\!R1$ diagram there is more scatter 
as the value for $H\!R1$ is less accurately determined than for $H\!R2$).

\section{Constraining intervening LMC gas columns}

In Paper\,I I have determined the $N^{\rm LMC}_{\rm HI}$ values by 
performing X-ray spectral fitting for individual AGN. Here a different
method, a hardness ratio analysis, has been chosen to constrain absorbing 
column densities. The chosen sample of AGN and candidate AGN comprises 
the AGN sample given in Paper\,I (22 AGN) and 64 additional candidate 
background X-ray sources. For 20 AGN and candidate AGN values (with 
1$\sigma$ errors) for the total LMC hydrogen absorbing column density 
$N^{\rm LMC}_{\rm H}$ could be derived and for additional 11 candidate 
AGN a range. For further 54 candidate AGN only 1$\sigma$ upper limits to 
the LMC gas column could be derived (and in addition in one case a 1$\sigma$ 
lower limit). It follows that the values derived for $N^{\rm LMC}_{\rm H}$ 
from the hardness ratio analysis are consistent with the values for the LMC 
absorbing column density derived from X-ray spectral fitting of Paper\,I.

I briefly outline the hardness ratio analysis method which has been applied.
I simulated powerlaw tracks in the $H\!R1$ -- $H\!R2$ plane for a wide range 
of powerlaw photon indices $-\Gamma = (0.8-3.0)$. I compared the location of 
the hardness ratio error ellipses of individual AGN and candidate AGN with 
respect to these tracks to infer the absorbing column densities of LMC gas in 
the direction of individual background X-ray sources. I assumed that the 
powerlaw photon indices of background sources are in the range 
$-\Gamma = (2.0-2.5)$, which is the range of powerlaw photon indices derived 
by Brinkmann et al. (2000) from the X-ray spectra of a large sample of AGN. 
In the simulations reduced metallicities ($-$0.3\,dex relative to galactic 
interstellar absorption abundances) have been assumed for the LMC gas and 
galactic interstellar abundances for the galactic foreground gas. In Tab.\,1 
the values for the galactic and LMC hydrogen column density 
$N^{\rm gal}_{\rm HI}$ and $N^{\rm LMC}_{\rm HI}$ are given which have been 
derived from 21-cm \HI\ surveys of the LMC field performed with the 
{\sl Parkes} radio telescope (Br\"uns et al 2001, see also Dickey \& Lockman 
1990). Gas columns derived from 21-cm measurements can be separated into a 
galactic and a LMC component due to the different systemic velocities of both 
components. In addition the total LMC absorbing hydrogen column density 
$N^{\rm LMC}_{\rm H}$ derived from the hardness ratio analysis is given.

 \begin{figure}[htbp]
  \centering{
  \vbox{\psfig{figure=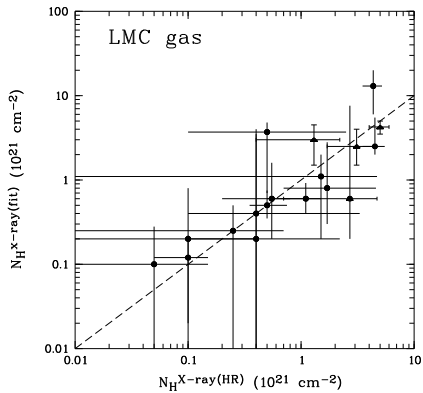,width=6.0cm,angle=0.0,%
  bbllx=1.5cm,bblly=1.0cm,bburx=19.2cm,bbury=17.0cm,clip=}}\par
            }
  \caption[]{Correlation between LMC hydrogen absorbing column density 
             (after galactic foreground gas has been removed) as derived 
             from the hardness ratio analysis compared with the LMC 
             hydrogen absorbing column density derived from the X-ray
             spectral fit (cf. Paper\,I). The dashed line gives the linear
             relation for which both LMC column density determinations 
             are equal.}
  \label{ps:nhcorr}
\end{figure}

In Fig.~{\ref{ps:nhcorr}} I show the correlation between the LMC hydrogen
absorbing column density derived with the hardness ratio analysis in
comparison with the LMC hydrogen absorbing column density derived from the
X-ray spectral fit (Paper\,I). There is a linear correlation between the LMC 
gas columns determined by both methods which gives reliability to the LMC 
gas columns derived by both methods.

 \begin{figure}[htbp]
  \centering{
  \vbox{\psfig{figure=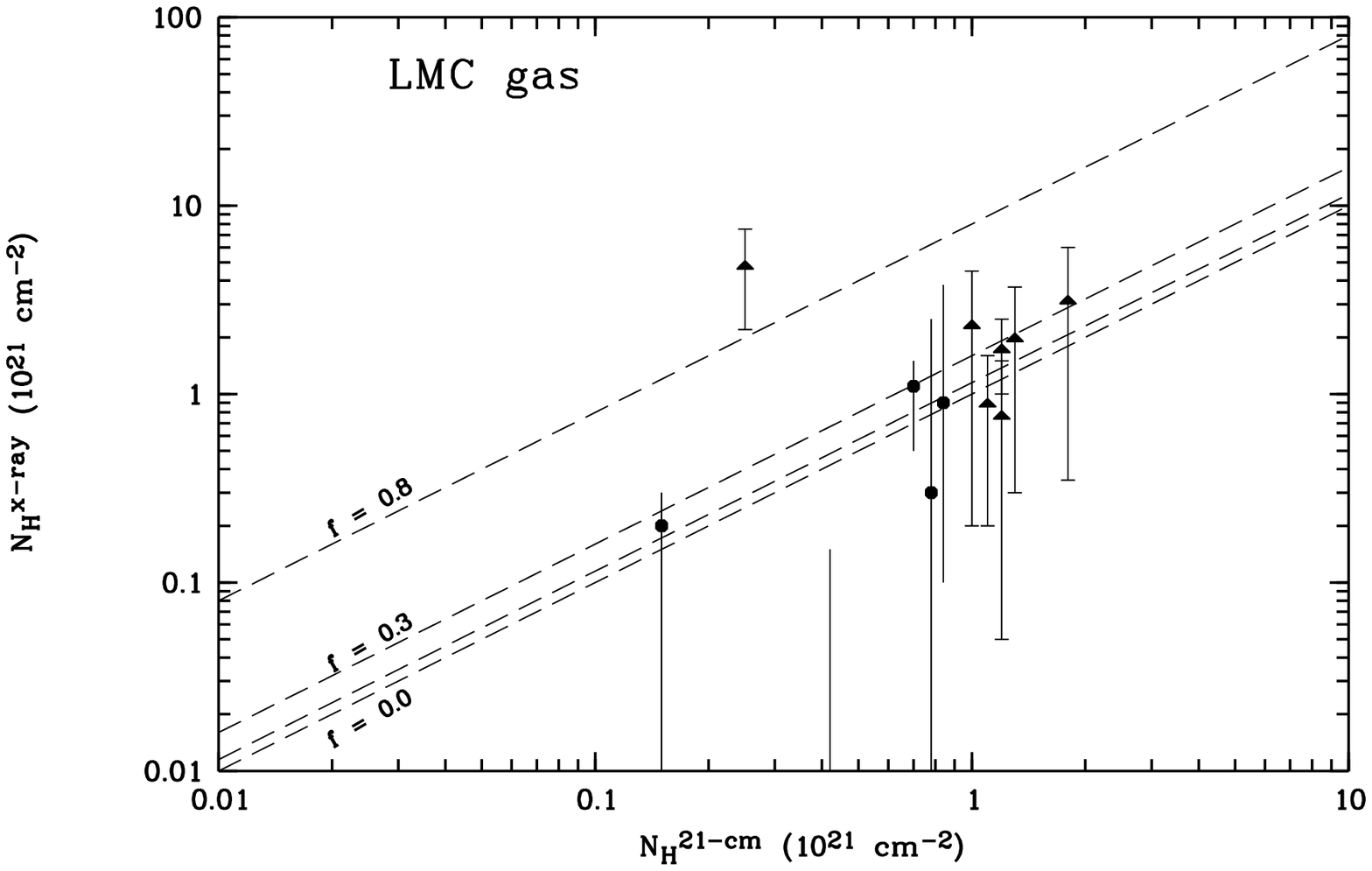,width=7.5cm,angle=0.0,%
  bbllx=1.5cm,bblly=1.5cm,bburx=20.5cm,bbury=13.5cm,clip=}}\par
  \vbox{\psfig{figure=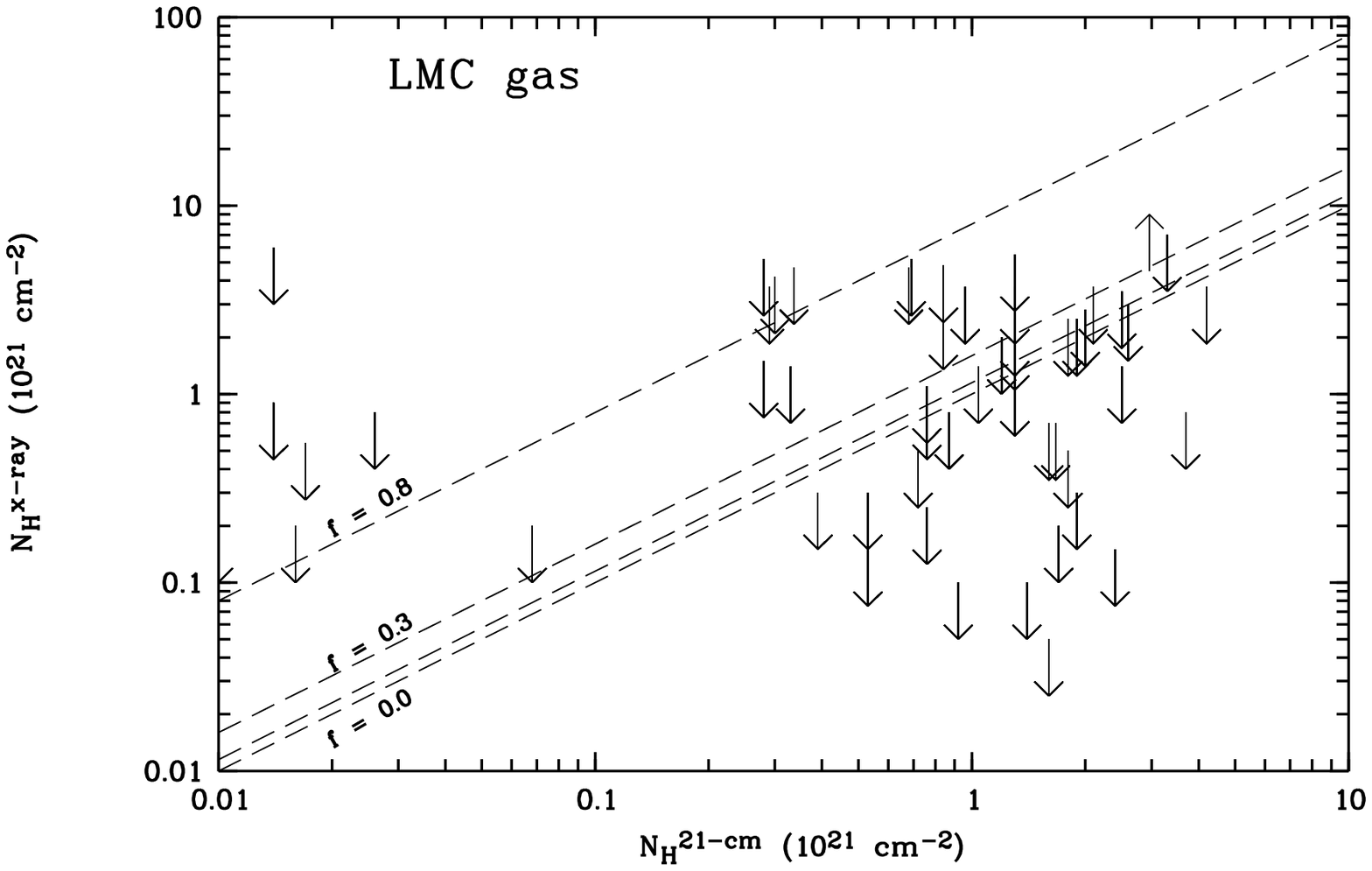,width=7.5cm,angle=0.0,%
  bbllx=1.5cm,bblly=1.5cm,bburx=20.5cm,bbury=13.5cm,clip=}}\par
            }
  \caption[]{LMC hydrogen absorbing column density (after galactic foreground
             gas has been removed) as derived from the hardness ratio
             analysis assuming constraints on the powerlaw photon index 
             (cf. Tab.\,1). Upper panel: AGN (additional to the sample 
             from Paper\,I) for which a best-fit has been determined (the 
             best-fit value is given as filled circle and 1$\sigma$ error 
             bars are drawn) and AGN for which an $N_{\rm H}$ range, band, 
             has been determined and for which the mean value is shown as a 
             filled triangle and 1$\sigma$ error bars are drawn with small 
             cross bars.
             Lower panel: AGN for which only upper limits (and in one case
             a lower limit) to the LMC gas columns has been determined. With 
             dashed lines the dependences on the molecular mass fraction 
             $f=0,\ 0.1,\ 0.3$ and $0.8$ are indicated.}
  \label{ps:lmcgas1}
\end{figure}

I present in Fig.~{\ref{ps:lmcgas1}} the comparison between the LMC gas 
columns inferred from the 21-cm \HI\ {\sl Parkes} survey and the LMC gas 
columns inferred from the hardness ratio analysis for background sources 
in addition to those for which X-ray spectral fitting has been performed 
in Paper\,I (for 6 AGN a best-fit for the LMC $N_{\rm H}$ value, for 7 AGN 
a range of values, for 50 AGN upper limits and for one AGN a lower limit is 
given). The values for the LMC gas columns inferred from the hardness ratio 
analysis agree in most cases within the uncertainties with the LMC \HI\ 
columns inferred from the {\sl Parkes} survey.

\subsection{Deriving constraints on the LMC metallicity}

It is assumed that AGN have canonical powerlaw photon indices $-\Gamma = 2.0$ 
to $2.5$ in the \ros\ {\sl PSPC}\ band. From simulations, tracks for constant 
powerlaw indices $-\Gamma$ and a galactic foreground absorbing column of
$3\ 10^{20}\ {\rm cm^{-2}}$ have been derived in the $H\!R1$ -- $H\!R2$ plane
by varying the LMC absorbing column density $N_{\rm H}$. 

For the LMC gas hybrid models have been used assuming a constant foreground 
hydrogen column due to the Milky Way gas of $3\times10^{20}\ {\rm cm}^{-2}$. 
In the simulations the metallicity has been varied from --0.8\,dex to 
+0.5\,dex in steps of 0.05\,dex. In addition $-\Gamma$ has been varied from
2.1 to 2.4 for the AGN and from 1.4 to 1.7 for the X-ray binaries. Three AGN 
(HP\,54, HP\,380 and HP\,1094, cf. Paper\,I) with accurately determined values
for the hardness ratios $H\!R1$ 
and $H\!R2$ could be used to constrain the metallicity. From the location of 
these AGN in the $H\!R1$ -- $H\!R2$ plane metallicities somewhat in excess 
of galactic metallicities $X>0.1$ can be excluded. Metallicities as low as 
$X$=--0.7 were found to be consistent with the data.

The size of the sample has been extended in a next step and the mean 
metallicity of the intervening LMC gas and the powerlaw slope of the flux 
have been determined in a least-square grid search. This search has been 
performed for two different samples, an AGN sample with 14 objects and an 
X-ray binary sample with 9 objects (cf. Tab.\,1). These objects were taken 
from a sample selected in this work and, in addition, only objects with 
accurate hardness ratios $\delta H\!R<0.20$ were used.

\begin{figure}
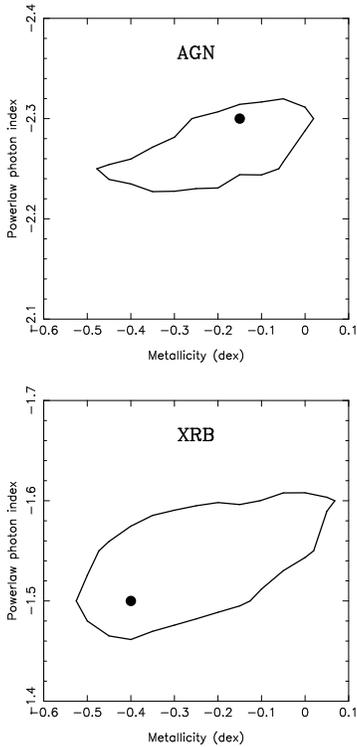
 
  \centering{
  \vbox{\psfig{figure=MS1832f8a.eps,width=5.0cm,angle=-90.0,%
  bbllx=4.0cm,bblly=4.2cm,bburx=19.0cm,bbury=19.0cm,clip=}}\par
  \vbox{\psfig{figure=MS1832f8b.eps,width=5.0cm,angle=-90.0,%
  bbllx=4.0cm,bblly=4.2cm,bburx=19.0cm,bbury=19.0cm,clip=}}\par
            }
  \caption[]{Upper panel: Confidence contours (99\%) for 14 candidate AGN 
  with catalog number 1, 37, 54, 101, 147, 380, 411, 561, 653, 876, 1040,
  1094, 1181, and 1247 (Haberl \& Pietsch 1999) and hardness ratio errors
  $\delta HR<0.20$) in the $\Gamma$ -- $X$ plane. 
  Lower panel: Confidence contours for 9 X-ray binaries (XRB) with catalog 
  number 41, 106, 184, 204, 252, 436, 1001, 1225, and 1325 and  
  $\delta HR<0.20$ in the $\Gamma$ -- $X$ plane.}
  \label{ps:figcon}
\end{figure}

From the formal fit it is found that the powerlaw slope $\Gamma $ and the 
metallicity $X$ can be constrained for both samples (the AGN and the X-ray
binary sample). In the case of the AGN, the errors in the hardness ratios
of the AGN have been increased by a factor of 1.4. The range of
metallicities which is derived in this way is in agreement with the range
of tracks for different metallicities which is covered by the used data
points.
In the case of the X-ray binaries, a systematic offset of 0.03 in the values 
of the hardness ratios has been assumed in the fit. This avoids 
that the least-square fit is biased towards the data points with very small 
error bars in the hardness ratio values as derived for the bright X-ray 
binaries (e.g. LMC~X-1).
This has the effect of increasing the parameter range for a given confidence 
(e.g. the 99\% confidence which is shown in Fig.~{\ref{ps:figcon}}).

In Fig.~{\ref{ps:figcon}} the confidence contours are shown for 
the 14 AGN and the 9 X-ray binaries in the $H\!R1$ -- $H\!R2$ plane. It is 
found that for the X-ray binaries the powerlaw slope can be confined to 
$-\Gamma$ = 1.45 to 1.6 and the metallicity to $X$ = --0.5 to +0.0 (99\% 
confidence). For the 14 AGN (Fig.~{\ref{ps:figcon}}) I find that the 
powerlaw slope can be confined to $-\Gamma$ = 2.2 to 2.3 and the metallicity 
to $X$ = --0.6 to +0.15 (99\% confidence). The best-fit metallicity is 
$-X = 0.4$ for X-ray binaries and $X = -0.15$ for AGN respectively.

The value for the metallicity which has been found from the AGN and the X-ray
binary sample is consistent with the metallicity of --0.2 to --0.6 derived 
for the LMC (cf. Pagel 1993; Russell \& Dopita 1992). The powerlaw photon 
index derived for the AGN sample of $-\Gamma$=2.2 to 2.3 is consistent with 
the powerlaw slope derived for AGN type spectra (cf. discussion in Paper\,I). 
The powerlaw photon index derived for the X-ray binaries $-\Gamma$=1.45 to 
1.6 is steeper than the canonical value of $-\Gamma \sim$1.0 (see also 
Sect.\,6). Apparently for the LMC X-ray binaries steeper powerlaw photon 
indices are observed in the \ros\ {\sl PSPC} band. From spectral fitting 
applied to bright LMC X-ray binaries (LMC~X-1, LMC~X-2, LMC~X-3 and LMC~X-4) 
follows that simple powerlaw spectra cannot explain the absorbed spectra and 
more complicated spectral shapes have to be fitted.

\section{The X-ray binary sample}

For the X-ray binary sample all classified X-ray binaries given in HP99 
have been considered. Three of these X-ray binaries were found to be slightly 
outside the hardness ratio selection criteria but were also included in the 
sample. In addition sources were taken into account which were observed in 
the central 20\arcmin of the detector, which fulfilled the selection 
criteria for X-ray binaries given in Sect.\,2, i.e. sources which were 
located in the $H\!R1$ -- $H\!R2$ plane ``above'' the AGN band (cf. Sect.\,4). 
In addition an X-ray spectral fit has been applied to the spectra of these 
sources and the consistency with an X-ray binary has been checked. Also a 
time variability study of the source count rate has been performed. There
were 30 sources found which were classified as (candidate) X-ray binaries 
(cf. Tab.\,1). 15 of these sources have more than 50 observed counts and 
powerlaw photon indices were derived for these sources (excluding HP\,914). 
It is found that the distribution of powerlaw photon indices is consistent 
with a mean $\Gamma$ of $-$1.4 and a $\sigma$ of 0.9. But $\Gamma$ strongly 
depends on the used value of the galactic and LMC absorbing column. For most 
of the X-ray binaries the total LMC columns have been used in the spectral 
fit. This assumption need not always be correct and I have taken this fact 
into account in a few cases where the $N_{\rm H}$ value could be determined 
in the spectral fit.

If one compares the number of classified AGN and X-ray binaries it is 
found that a fraction of 80\% of the spectrally hard X-ray sources with more 
than 50 detected counts are AGN and 20\% are X-ray binaries. 

\begin{figure} 
  \centering{
  \vbox{\psfig{figure=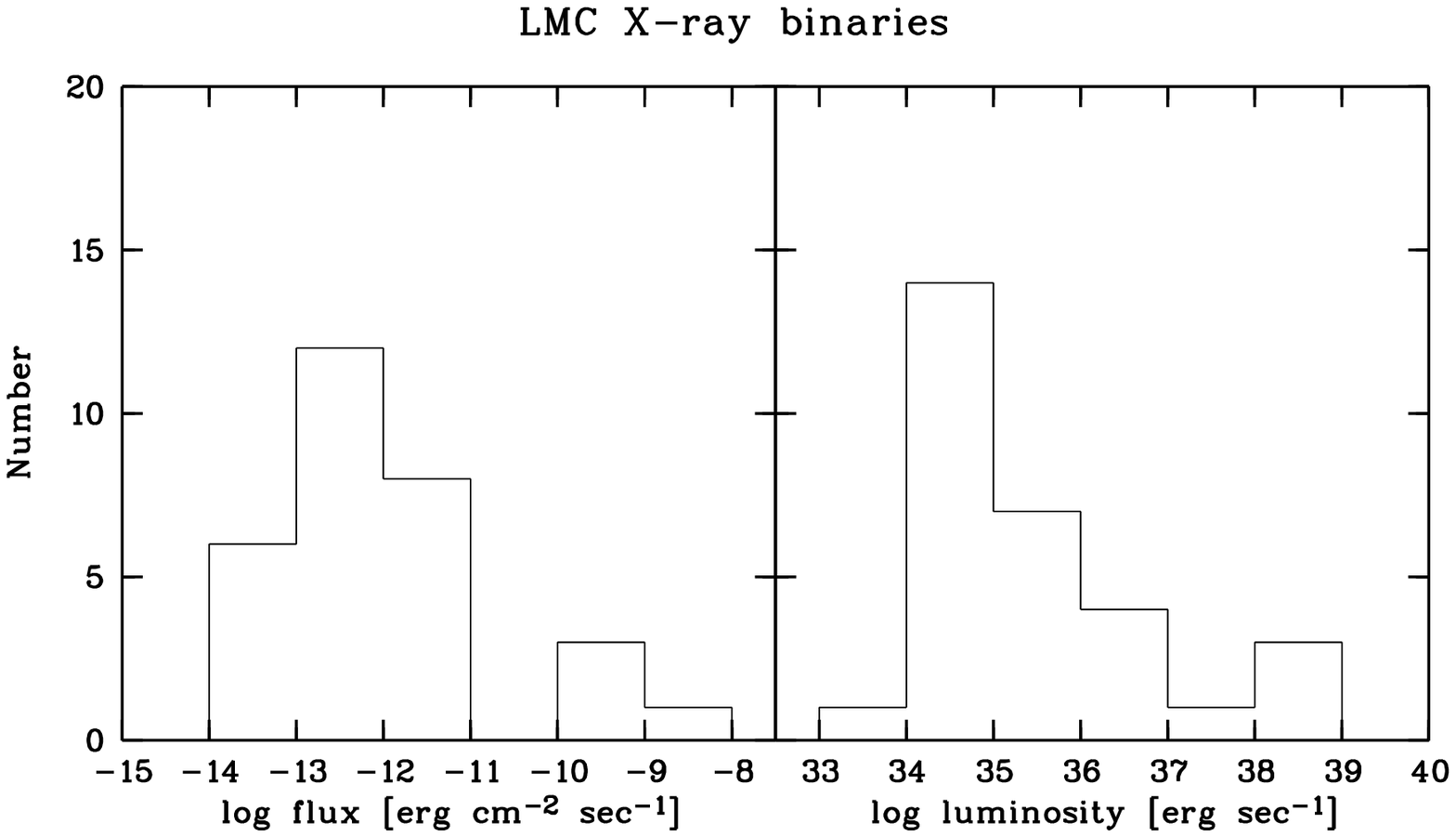,width=8.0cm,angle=0.0,%
  bbllx=1.0cm,bblly=1.5cm,bburx=19.0cm,bbury=11.2cm,clip=}}\par
            }
  \caption[]{Distribution of flux and luminosity corrected for absorption
  (left and right panel respectively) for the 30 (candidate) X-ray binaries 
  in the observed field of the LMC (cf. Tab.\,1). The number of X-ray 
  binaries per flux and luminosity bin is given.} 
  \label{ps:xrbfl}
\end{figure}

For the 30 (candidate) X-ray binaries the unabsorbed flux and the luminosity 
(0.1-2.4)~keV have been determined in an X-ray spectral fit. The derived
flux and luminosity histograms are given in Fig.~{\ref{ps:xrbfl}}. It follows
that there are 3, 4, 8, 15, and 29 X-ray binaries with luminosities in excess 
of $10^{38}$, $10^{37}$, $10^{36}$, $10^{35}$, and 
$10^{34}\ {\rm erg}\ {\rm s}^{-1}$ respectively.

These numbers can be compared with the number of X-ray binaries predicted
from stellar evolutionary calculations for the LMC (Dalton \& Sarazin, 1995).
According to these calculations there are 1, 5, 18, 125, and 750 X-ray
binaries predicted to exist in the LMC with luminosities in excess of
$10^{38}$, $10^{37}$, $10^{36}$, $10^{35}$, and
$10^{34}\ {\rm erg}\ {\rm s}^{-1}$ respectively. Such a comparison will only 
be valid if the sample of X-ray binaries selected here is complete. There
are two factors which have to be taken into account for such a completeness
consideration, the sensitivity limit of the LMC X-ray survey and the 
fraction of the LMC disk covered by the observations.

In Paper\,III it will
be shown that our survey is complete in the observed field to a flux of
$\sim$$10^{-12}\ {\rm erg}\ {\rm cm}^{-2}\ {\rm s}^{-1}$ which is equivalent
to a luminosity of $3.\ 10^{35}\ {\rm erg}\ {\rm s}^{-1}$. Our observations 
cover 16 square degrees, which is $\sim$24\% of the LMC disk. Assuming
that X-ray binaries are homogenously distributed across the LMC disk, we 
extrapolate from the number of 8 observed X-ray binaries with luminosities in 
excess of $10^{36}\ {\rm erg}\ {\rm s}^{-1}$ that there may be 33 X-ray 
binaries across the whole LMC disk above this luminosity limit. If one 
compares this with the number of 18 X-ray binaries predicted from population
synthesis calculations then an excess of X-ray binaries appears to exist.
But a detailed investigation of the candidate X-ray binaries is required to
give reliability to a deviation in these numbers. For a flux in excess of
$3.\ 10^{-13}\ {\rm erg}\ {\rm cm}^{-2}\ {\rm s}^{-1}$ (which corresponds 
to a luminosity of $10^{35}\ {\rm erg}\ {\rm s}^{-1}$) 15 X-ray binaries are
in our observed sample. Assuming besides the incompleteness due to the covered
LMC field the incompleteness due to the given sensitivity (which is about a
factor of 1.3, cf. Paper\,III) we derive an extrapolated population of 81 
X-ray binaries. This population would be less than the predicted 125 X-ray 
binaries but the extrapolated number my have large uncertainties and such a 
comparison may not be too reliable.

\subsection{Comparison with the number of X-ray binaries in the SMC} 

Two of the newly classified X-ray binaries, RX~J0523.2-7004 and 
RX~J0527.1-7005, are located in the optical bar of the LMC (cf. Tab.\,1). 
Another source newly classified as an X-ray binary, RX~J0524.2-6620, lies in 
the eastern \HI\ shell of the supergiant shell LMC~4. An additional source 
which is contained in Tab.\,1 in the section of background AGN, 
RX~J00546.8-6851, but which may be an X-ray binary (see also Paper\,I and 
Sasaki, Haberl, \& Pietsch 2000) is located in or at least very close to the 
supergiant shell LMC~2. In total, 9 of the 30 sources classified as X-ray
binaries (i.e. 30\%) are associated with the supergiant shell LMC~4. This
could be a selection effect as the LMC~4 region has been observed during
many {\sl ROSAT} pointings. But also other regions of the LMC, e.g. the
30~Dor area, have been observed during multiple observations and  less
X-ray binaries have been detected in these areas. Assuming that these 
sources are high-mass X-ray binary systems which have formed within an 
evolutionary time scale of $\sim$$10^{7}\ {\rm years}$ (cf. Popov et al. 
1998) may indicate that star formation has taken place in the last 10 
million years in the LMC disk (including the \HI\  boundary of the 
supergiant shell LMC~4). To find candidate high-mass X-ray binaries in the 
LMC may be of importance as recent X-ray surveys of the Small Magellanic 
Cloud (SMC) have revealed a large number of such systems showing 
X-ray pulsations in this other Magellanic Clouds galaxy
(cf. Yokogawa et al. 2000; Finger et al. 2001). One scenario put forward 
to explain the large number of high-mass X-ray binaries discovered in the 
SMC is the trigger of star formation during the recent close encounter 
between the SMC and the LMC $\sim$(0.2--0.4)~Gyr ago (cf. Gardiner et al. 
(1994, hereafter GSF94); Gardiner \& Noguchi (1996, hereafter GN96)). In 
such a scenario it is expected that star formation was also triggered in 
the LMC (cf. van den Bergh, 2000, for a recent update of the star formation
rate of the LMC during the last 9~Gyr). Finding new candidate high-mass 
X-ray binaries in the LMC which are associated with at least two supergiant 
shells may be consistent with such a scenario.

Can this scenario account for the observed number of candidate high-mass 
X-ray binaries in the LMC and SMC. In the previous section we estimated an 
extrapolated number of 33 X-ray binaries with luminosities in excess of 
$10^{36}\ {\rm erg}\ {\rm s}^{-1}$ in the LMC field. A comparable number  
for the population of high-mass X-ray binaries in the SMC has been set up 
by Haberl \& Sasaki (2000) who recently increased the number of detected 
Be-type X-ray binaries in the SMC to $\sim$50. Assuming that at least 40\% 
of these X-ray binaries have outburst luminosities in excess of 
$10^{36}\ {\rm erg}\ {\rm s}^{-1}$ would give a ratio of LMC to SMC high-mass 
X-ray binaries of $\sim$(0.7--1.7). An additional uncertainty in these 
numbers may be due to the fact that not all Be-type X-ray binaries have
so far been detected in the LMC and the SMC (either in quiescence or in
outburst). A value for the number ratio of $\sim$(0.7--1.7) is not in
agreement with the mass ratio of both galaxies of $\sim$10 (the mass of the LMC
and the SMC is $\sim$$2\ 10^{10}\ M_{\odot}$ and $\sim$$2\ 10^{9}\ M_{\odot}$ 
respectively, cf. GSF94). It would be more consistent with the ratio of the 
gas mass of both galaxies of $\sim$(1.2--1.8) (the \HI\ mass of the LMC and SMC is 
$\sim$$5.2\ 10^{8}\ M_{\odot}$ (Kim et al. 1998) and $\sim$$4.2\ 10^{8}\ M_{\odot}$ 
(Stanimirovic et al. 1999) respectively, and for the LMC the gas mass may be 
larger than the \HI\ mass by $\sim$40\% due to the contribution of molecular 
hydrogen). Assuming that the star formation rate is proportional to the gas 
mass of a galaxy, the comparable gas mass of the SMC and the LMC may give an 
explanation for the comparable number of high-mass X-ray binaries found in 
both galaxies. 

Star formation may have been triggered during an encounter of these two
galaxies. Assuming that during the encounter turbulence was introduced into 
the gaseous phase of the galaxy disk, from the condition of conservation of 
angular momentum constraints can be derived for the ratio of star formation 
rates $SFR$ induced in both galaxies. Making use of the formalism for the
star formation rate given by Kennicutt (1998) in which the star formation 
rate scales with the gas density and the orbital time scale and which has 
been found to give a good fit for a large sample of normal and starburst 
galaxies, then one finds that this ratio can be expressed\footnote{Eq.\,3 
has been derived assuming conservation of angular momentum during the 
encounter of the LMC and the SMC. As a result the SMC disk was spun up to 
a higher angular velocity than the LMC disk (but see Maragoudaki et al. 
2001). This resulted in a more efficient star formation burst in the SMC 
compared to the LMC. The history of the interaction, i.e. the state of the 
system prior to the encounter, and the distance of both galaxies during the 
encounter do not enter into this consideration. Only the ratio of the star 
formation rates induced by the encounter of both galaxies is given. Only 
under the assumption that the encounter was close enough to enhance the 
star formation rate efficiently Eq.\,3 may be of relevance for the 
consideration of the total star formation rate during the encounter. 
Different values have been derived from N-body and SPH simulations of 
such a galaxy-galaxy encounter for the distance of closest approach, 7~kpc, 
(GSF94, GN96), and  20~kpc, (Li \& Thronson 1999), respectively. Any 
determination of the absolute value of the induced star formation rate 
depends on the specific assumptions made about the history of this galaxy 
encounter.}   

as

\begin{equation}  
  \frac{SFR_{\rm LMC}}{SFR_{\rm SMC}} \approx 
  \big(\frac{R_{\rm SMC}}{R_{\rm LMC}}\big)^2
\end{equation}

with $R$ the radius of the gaseous disk of a galaxy. If one uses for the LMC
$R_{\rm LMC}$=3.7~kpc (Kim et al. 1998) and for the SMC $R_{\rm SMC}$=2.3~kpc 
(e.g. Stanimirovic et al. 1999), then one obtains 
$\frac{SFR_{\rm LMC}}{SFR_{\rm SMC}}\approx 0.4$

If one assumes that the starburst was efficient enough to significantly 
increase the star formation rate preferentially in the SMC and that the 
number of high-mass X-ray binaries scales with the star formation rate of 
a galaxy at an epoch of $\sim$$10^{7}$ years ago (which may be somewhat 
earlier if a delay for the onset of star formation is taken into account) 
then one can directly compare the ratio of the star formation rates of two 
galaxies during this epoch with the ratio of presently observed numbers of 
high-mass X-ray binaries in these galaxies. The ratio of high-mass X-ray 
binaries in the LMC to those in the SMC is derived from the 
observed numbers to be $\sim$(0.7-1.7). There appear to be many more 
X-ray binaries in the LMC than predicted from Eq.\,3. One explanation 
may be that the formation of high-mass X-ray binaries in the LMC is less 
affected by the starburst than in the SMC, i.e. in the LMC we observe the 
constant star formation with a minor contribution from a starburst.

From the {\sl OGLE} survey of 93 star clusters in a field in the central 
$2.4\ {\rm deg}^{2}$ of the SMC Pietrzynski \& Udalski (1999) derived that 
most of these star clusters are younger than $\sim$$20.\ 10^{7}\ {\rm years}$. 
This finding could mean that the formation of star clusters during the last 
$(20 - 30)\ 10^{7}\ {\rm years}$ was enhanced at least in the central field
of the LMC. Alternatively it may be explained by an efficient process of
disintegration of clusters older than $(20 - 30)\ 10^{7}\ {\rm years}$.
Both effects may be explained by a tidal interaction of the SMC with the
LMC which may have resulted in a burst of cluster formation and/or in the
disruption of pre-existing stellar clusters.

\section{Summary and conclusions}

The sample of spectrally hard X-ray sources in the field of the LMC observed 
with the \ros\ {\sl PSPC} and published in the catalog of HP99 has been
reinvestigated. Especially accurate values for the count rate have been
determined in the spectrally hard (0.5 -- 2.0\,keV) and broad (0.1 -- 2.4\,keV)
band respectively and values for the hardness ratios  $H\!R1$ 
and $H\!R2$ have been determined making use of merged data in the direction 
of each investigated X-ray source. The analysis has been restricted to X-ray 
sources which have been observed in the inner 20\arcmin\ of the {\sl PSPC} 
detector.

Simulations have been performed to derive tracks for powerlaw spectra with 
slopes comprised by X-ray binaries and AGN in the $H\!R1$ -- $H\!R2$ plane. 
In these simulations a wide range of metallicities for the LMC gas has been 
considered. Comparing the location of the X-ray sources in the 
$H\!R1$ -- $H\!R2$ plane with respect to the simulated tracks for X-ray 
binaries and AGN a source classification has been achieved of the sample of 
spectrally hard X-ray sources observed in the central 20\arcmin\ of the 
{\sl PSPC} detector. 141 sources have been classified as AGN (or as likely 
AGN) and 30 sources as X-ray binaries (or as likely XRB). This means that 
82\% of the classified hard X-ray sources in the LMC field are AGN and 18\% 
are X-ray binaries.

I constrained, for 31 of these AGN (18 in addition to sources already 
investigated in Paper\,I), the LMC gas columns from the location of these 
sources in the $H\!R1$ -- $H\!R2$ plane. In addition I derived for 54 AGN 
upper limits for the LMC gas columns.

I independently constrained the metallicity of the LMC gas by fitting 
simulated tracks of constant powerlaw slopes in the $H\!R1$ -- $H\!R2$ plane 
to the observation derived $H\!R1$ and $H\!R2$ values for the AGN and the 
X-ray binary sample. I found that the required metallicity of the LMC gas is 
in the range $-$0.6 to $+$0.1\,dex at 99\% confidence. 

I also established the catalog of X-ray sources in a deep merged observation
of the field of the supergiant shell LMC\,4. I detected 97 X-ray sources
of which I classified 35 sources as candidate AGN.

The number of 30 (candidate) X-ray binaries observed in the LMC is compared 
with the number of X-ray binaries predicted from population synthesis 
calculations for the LMC. In addition the number of (candidate) high-mass 
X-ray binaries observed in the LMC with luminosities in excess of 
$10^{36}\ {\rm erg}\ {\rm s}^{-1}$ is compared with the number of high-mass 
X-ray binaries in the SMC. It is found that the comparable number of high-mass
X-ray binaries scale with the comparable gas mass of the host galaxies.
The number of high-mass X-ray binaries may have been preferentially 
enhanced in the SMC due to a star formation burst initiated by the LMC-SMC
galaxy encounter.

\appendix

\section{X-ray sources in a deep merged observation of the northern field
of the supergiant shell LMC\,4}

In the previous analysis in this paper I have made use of the X-ray sources 
given in the catalog of HP99. Part of the LMC region has been observed during 
several observations and a large integrated exposure exists for these fields. 
In particular I investigated the field of the northern area of the supergiant 
shell LMC\,4 to find out how many X-ray sources can be detected in deep merged 
observations of this specific field.

LMC\,4 is the northern and largest (with a diameter of $\sim$1200~pc) of 
five supergiant shells in the LMC which are characterized by circular regions 
of filamentary \HII\ emission (Meaburn 1980). McGee \& Milton (1966) noted
the existence of a deep minimum in the column density of neutral hydrogen 
in LMC\,4 in their 21-cm \HI\ observations. Inside this hole is the stellar
association Shapley~III (Nail \& Shapley 1953).

I merged the observations\footnote{The center of the merged observation 
is R.A. 05$^{\rm h}$29$^{\rm m}$00$^{\rm s}$, Decl. 
-66$^{o}$03$\arcmin$04${\arcsec}$}
existing for the field in the northern area of the supergiant shell LMC\,4
making use of source detection routines (local, map, and maximum likelihood) 
which are available in {\sl EXSAS}. I applied the local, map, and the 
maximum likelihood source detection task to one energy band (0.5 -- 2.4~keV)
to derive the catalog of sources. 97 X-ray sources were detected in the 1.8 
square degree field. I accepted only sources with a likelihood ratio of 
existence $>8$ and checked the reality of the sources on a displayed image. 
The catalog of these sources is given in Tab.\,3. Column~1 gives the source 
index, Column~2 the \ros\ source name, Column~3 and 4 the source position, 
the right ascension (RA) and the declination (Dec) for the epoch J2000 with 
the 90\% confidence positional uncertainty (Column~5), the count rate 
(0.1 -- 2.4~keV, Column~6), the hardness ratios $H\!R1$ and $H\!R2$ (Column~7 
and 8), the source extent in arcsec in case the extent likelihood ratio is 
$>10$ (Column~9), the likelihood ratio of existence $L_{\rm exi} = - ln(P)$, 
with $P$ the probability that the detected source is due to excess counts 
measured above a mean local background (Column~10), the distance of the source 
from the center of the field of the merged observation (Column~11), the source 
index from the catalog of HP99 (Column~12), the distance to that source in 
arcsec (Column~13), the classification of the source (with A = AGN, B = X-ray 
binary, R = supernova remnant, F = foreground star) in Column~14 and remarks 
in Column~15.

The exposure time of the merged observation varies over a large range 
and has a region of high exposure ($\sim50$ to $70$~ksec) in one merged 
{\sl PSPC} pointing. The limiting flux for the sources detected in this 
merged observation and given in the catalog of Tab.\,3 is 
$8\ 10^{-15}\ {\rm erg}\ {\rm cm^{-2}}\ {\rm s^{-1}}$ assuming a powerlaw 
spectrum with a photon index of $-\Gamma$=2 and a galactic absorbing column 
density of $5\ 10^{20}\ {\rm cm^{-2}}$.

If one considers the location of the sources detected in this field with 
accurate values for the hardness ratios $\delta H\!R1 \le 0.20$ and 
$\delta H\!R2 \le 0.20$ in the hardness ratio $H\!R1$ -- $H\!R2$ plane
then one finds that most of the sources are located in the region bounded 
by the powerlaw tracks $-\Gamma$~=~1.0, 2.0 and 3.0 and are consistent 
with AGN and X-ray binaries (cf. Sect.\,4). A few bright sources which are 
located outside this region are (consistent with) supernova remnants.

A fraction of the detected sources is contained in the sample of sources 
investigated in the previous sections and is also given in Tab.\,1. 60 
sources correlate within a search radius of 20\arcsec\ with a source in the 
catalog of Haberl \& Pietsch (1999). I made a tentative classification of 
the 97 X-ray sources detected in the field of the supergiant shell LMC\,4 
from the location of these sources in the $H\!R1$ - $H\!R2$ plane 
(cf. Sect.\,4.1). I classified 35 of the sources with more than 30 
observed counts as AGN (or likely AGN). This low threshold in counts 
has been chosen as the X-ray survey in the field of the Supergiant Shell
LMC\,4 was considerably deeper than the X-ray survey of the general LMC 
field. In Paper\,III it will be made use of this AGN sample to derive the 
$\log N - \log S$ of background X-ray sources in the field of the Supergiant
Shell LMC\,4.

\acknowledgements
%________________________________________ Do not leave a blank line here!
The \ros\ project is supported by the Max-Planck-Gesellschaft and the 
Bundesministerium f\"ur Forschung und Technologie (BMFT). This research has 
made use of the {\sl SIMBAD} data base operated at CDS, Strasbourg, France.
I thank C. Br\"uns for making available the {\sl Parkes} 21-cm map of the 
galactic and LMC \HI\ in the field of the LMC. I thank J. Kerp for comments 
on an earlier version of the manuscript. I thank K.S. de Boer for suggestions 
to improve the article. I thank an anonymous referee for the suggestions to
improve the manuscript. PK is supported by the Graduiertenkolleg on the 
``Magellanic Clouds and other Dwarf galaxies'' (DFG GRK\,118).
%_____________________________________________________________________

\end{document}